\documentclass[10pt]{article}
\usepackage{graphicx}
\usepackage{hyperref}
\def\@magscale#1{ scaled \magstep #1}
\def\fracm#1#2{\hbox{\large{${\frac{{#1}}{{#2}}}$}}}

\def\stackunder#1#2{\mathrel{\mathop{#2}\limits_{#1}}}%

\catcode`@=11  
\def\un#1{\relax\ifmmode\@@underline#1\else
        $\@@underline{\hbox{#1}}$\relax\fi}
\catcode`@=12

\def\a{\alpha}
\def\b{\beta}

\def\d{\delta}
\def\e{\epsilon}

\def\m{\mu}
\def\n{\nu}
\def\o{\omega}

\def\r{\rho}
\def\s{\sigma}
\def\t{\tau}
\def\u{\upsilon}
\def\x{\xi}
\def\z{\zeta}

\def\G{\Gamma}

\def\L{\Lambda}
\def\O{\Omega}

\def\dslash{\not{\hbox{\kern-2pt $\partial$}}}
\def\Dslash{\not{\hbox{\kern-4pt $D$}}}
\def\pslash{\not{\hbox{\kern-2.3pt $p$}}}
 \newtoks\slashfraction
 \slashfraction={.13}
 \def\slash#1{\setbox0\hbox{$ #1 $}
 \setbox0\hbox to \the\slashfraction\wd0{\hss \box0}/\box0 }
 
\font\ro=cmsy10                          % font with rope
\def\kcr{{\hbox{\ro \char'170}}}                % right-handed rope
\def\ktl{{\hbox{\ro \char'170}}}        % top end for left-handed rope
\def\ktr{{\hbox{\ro \char'170}}}        % " right
\def\kbl{{\hbox{\ro \char'170}}}        % " bottom left
\def\kbr{{\hbox{\ro \char'170}}}        % " right
\def\plpl{\raise-2pt\hbox{$\raise3pt\hbox{$_+$}\hskip-6.67pt\raise0.0pt
\hbox{$^+$}\hskip 0.01pt$}}
\def\mimi{\raise-2pt\hbox{$\raise3pt\hbox{$_-$}\hskip-6.67pt\raise0.0pt
\hbox{$^-$}\hskip 0.01pt$}} 

\def\bo{{\raise.15ex\hbox{\large$\Box$}}}               % D'Alembertian
\def\pa{\partial}                                       % curly d
\def\TH{{\raise.2ex\hbox{$\displaystyle \bigodot$}\mskip-4.7mu \llap H\;}}
\def\face{{\raise.2ex\hbox{$\displaystyle \bigodot$}\mskip-2.2mu \llap{$\ddot \smile$}}}                                      % happy face
\def\leftrightarrowfill{$\mathsurround=0pt \mathord\leftarrow \mkern-6mu
        \cleaders\hbox{$\mkern-2mu \mathord- \mkern-2mu$}\hfill
        \mkern-6mu \mathord\rightarrow$}
\def\dvec#1{\vbox{\ialign{##\crcr
        \leftrightarrowfill\crcr\noalign{\kern-1pt\nointerlineskip}
        $\hfil\displaystyle{#1}\hfil$\crcr}}}           % <--> accent
\def\fracm#1#2{\hbox{\large{${\frac{{#1}}{{#2}}}$}}}
\def\frac#1#2{{\textstyle{#1\over\vphantom2\smash{\raise.20ex
        \hbox{$\scriptstyle{#2}$}}}}}                   % fraction
\def\sfrac#1#2{{\vphantom1\smash{\lower.5ex\hbox{\small$#1$}}\over
        \vphantom1\smash{\raise.4ex\hbox{\small$#2$}}}} % alternate fraction
\def\bfrac#1#2{{\vphantom1\smash{\lower.5ex\hbox{$#1$}}\over
        \vphantom1\smash{\raise.3ex\hbox{$#2$}}}}       % "
\def\afrac#1#2{{\vphantom1\smash{\lower.5ex\hbox{$#1$}}\over#2}}    % "
\newskip\humongous \humongous=0pt plus 1000pt minus 1000pt

\newif\ifdtup

\parskip=\medskipamount                 % space between paragraphs (LaTeX)
\thicklines                         % thick straight lines for
\def\oldheadpic{                                % old UM heading
        \setlength{\unitlength}{.4mm}
        \thinlines
        \par
        \begin{picture}(349,16)
        \put(325,16){\line(1,0){4}}
        \put(330,16){\line(1,0){4}}
        \put(340,16){\line(1,0){4}}
        \put(335,0){\line(1,0){4}}
        \put(340,0){\line(1,0){4}}
        \put(345,0){\line(1,0){4}}
        \put(329,0){\line(0,1){16}}
        \put(330,0){\line(0,1){16}}
        \put(339,0){\line(0,1){16}}
        \put(340,0){\line(0,1){16}}
        \put(344,0){\line(0,1){16}}
        \put(345,0){\line(0,1){16}}
        \put(329,16){\oval(8,32)[bl]}
        \put(330,16){\oval(8,32)[br]}
        \put(339,0){\oval(8,32)[tl]}
        \put(345,0){\oval(8,32)[tr]}
        \end{picture}
        \par
        \thicklines
        \vskip.2in}
\def\oldtitle#1#2#3#4{\oldheadpic\begin{center}\vglue.5in{\large\bf
#1}\\[.6in]
        {#2}\\[.1in] {\it Department of Physics and Astronomy}\\
        {\it University of Maryland, College Park, MD 20742}\\[.6in]
        Physics Publication \#{#3}\\ {#4}\\[1.5in] {\bf
ABSTRACT}\\[.1in]
        \end{center} \begin{quotation}}                 % old title stuff
\def\oldTitle#1#2#3#4#5#6#7{\oldheadpic\begin{center} \vglue .4in
        {\large\bf #1}\\[.4in]
        {#2}\\[.1in] {\it Department of Physics and Astronomy}\\
        {\it University of Maryland, College Park, MD 20742}\\[.1in]
        {#3}\\[.1in] {\it {#4}}\\ {\it {#5}}\\[.4in]
        Physics Publication \#{#6}\\ {#7}\\[.5in] {\bf ABSTRACT}\\[.1in]
        \end{center} \begin{quotation}}                 % " for 2 authors
\def\border{                                            % border
        \setlength{\unitlength}{1mm}
        \newcount\xco
        \newcount\yco
        \xco=-21
        \yco=12
        \begin{picture}(140,0)
        \put(\xco,\yco){$\ktl$}
        \advance\yco by-1
        {\loop
        \put(\xco,\yco){$\kcr$}
        \advance\yco by-2
        \ifnum\yco>-240
        \repeat
        \put(\xco,\yco){$\kbl$}}
        \xco=158
        \yco=12
        \put(\xco,\yco){$\ktr$}
        \advance\yco by-1
        {\loop
        \put(\xco,\yco){$\kcr$}
        \advance\yco by-2
        \ifnum\yco>-240
        \repeat
        \put(\xco,\yco){$\kbr$}}
        \put(-20,13){\tiny ***University of Maryland **** Center for String and 
         Particle  Theory**** Physics Department****University of Maryland ****Center  
        for String and Particle  Theory** }
        \put(-20,-241.5){\tiny The University of Iowa Nuclear and Particle Theory
          Group The University of Iowa Nuclear and Particle Theory
          Group The University of Iowa  Nuclear and Particle Theo}
        \end{picture}
        \par\vskip-8mm}
\def\bordero{                                           % alternateborder
        \setlength{\unitlength}{1mm}
        \newcount\xco
        \newcount\yco
        \xco=-31
        \yco=12
        \begin{picture}(140,0)
        \put(\xco,\yco){$\ktl$}
        \advance\yco by-1
        {\loop
        \put(\xco,\yco){$\kclr$}
        \advance\yco by-2
        \ifnum\yco>-240
        \repeat
        \put(\xco,\yco){$\kbl$}}
        \xco=151
        \yco=12
        \put(\xco,\yco){$\ktr$}
        \advance\yco by-1
        {\loop
        \put(\xco,\yco){$\kcr$}
        \advance\yco by-2
        \ifnum\yco>-240
        \repeat
        \put(\xco,\yco){$\kbr$}}
        \put(-20,12){\ooobacdefghidfghghdhededbihdgdfdfhhdheidhdheb
         aaahjhhdahbahgdedgehgfdiehhgdigicba}
        \put(-20,-241.5){\oooababaighefdbfghgeahgdfgafagihdidihiidhiag
        fedhadbfdecdcdfagdcbhaddhbgfchbgfdacfediacbabab}
        \end{picture}
        \par\vskip-8mm}
\def\headpic{                                           % UM heading
        \indent
        \setlength{\unitlength}{.4mm}
        \thinlines
        \par
        \begin{picture}(29,16)
        \put(165,16){\line(1,0){4}}
        \put(170,16){\line(1,0){4}}
        \put(180,16){\line(1,0){4}}
        \put(175,0){\line(1,0){4}}
        \put(180,0){\line(1,0){4}}
        \put(185,0){\line(1,0){4}}
        \put(169,0){\line(0,1){16}}
        \put(170,0){\line(0,1){16}}
        \put(179,0){\line(0,1){16}}
        \put(180,0){\line(0,1){16}}
        \put(184,0){\line(0,1){16}}
        \put(185,0){\line(0,1){16}}
        \put(169,16){\oval(8,32)[bl]}
        \put(170,16){\oval(8,32)[br]}
        \put(179,0){\oval(8,32)[tl]}
        \put(185,0){\oval(8,32)[tr]}
        \end{picture}
        \par\vskip-6.5mm
        \thicklines}
\def\title#1#2#3#4{\border\headpic {\hbox to\hsize{#4 \hfill UMDEPP #3}}\par
        \begin{center} \vglue .5in {\large\bf #1}\\[.6in]
        {#2}\\[.1in] {\it Department of Physics and Astronomy}\\
        {\it University of Maryland, College Park, MD 20742}\\[1.5in]
        {\bf ABSTRACT}\\[.1in] \end{center} \begin{quotation}}  % title stuff
\def\Title#1#2#3#4#5#6#7{\border\headpic
        {\hbox to\hsize{#7 \hfill UMDEPP #6}}\par
        \begin{center} \vglue .4in {\large\bf #1}\\[.4in]
        {#2}\\[.1in] {\it Department of Physics and Astronomy}\\
        {\it University of Maryland, College Park, MD 20742}\\[.1in]
        {#3}\\[.1in] {\it {#4}}\\ {\it {#5}}\\[.5in] {\bf ABSTRACT}\\[.1in]
        \end{center} \begin{quotation}}                 % " for 2 authors
\def\endtitle{\end{quotation}\newpage}                  % end title page

\def\ad{{\kern0.5pt
                   \alpha \kern-5.05pt
\raise5.8pt\hbox{$\textstyle.$}\kern 0.5pt}}

\def\bd{{\kern0.5pt
                   \beta \kern-5.05pt
\raise5.8pt\hbox{$\textstyle.$}\kern 0.5pt}}

\def\qd{{\kern0.5pt
                   q \kern-5.05pt \raise5.8pt\hbox{$\textstyle.$}\kern0.5pt}}
\def\Dot#1{{\kern0.5pt
                   {#1} \kern-5.05pt
\raise5.8pt\hbox{$\textstyle.$}\kern0.5pt}}

\begin{document}

\def\gfrac#1#2{\frac {\scriptstyle{#1}}
        {\mbox{\raisebox{-.6ex}{$\scriptstyle{#2}$}}}}
\def\gg{{\hbox{\sc g}}}
\border\headpic {\hbox to\hsize{March 2003 \hfill
{UMDEPP 03-025}}}
\par
\par
\setlength{\oddsidemargin}{0.3in}
\setlength{\evensidemargin}{-0.3in}
\begin{center}
\vglue .10in
{\large\bf Short Distance Expansion from the \\Dual Representation of\\
  Infinite Dimensional Lie Algebras
\footnote {Supported 
in part by National  Science Foundation Grant PHY-0099544}  }
\\[.5in]
S. James Gates, Jr.\footnote{gates@physics.umd.edu}, W.D. Linch,
III \footnote{ldw@physics.umd.edu}, Joseph Phillips \footnote{ferrigno@physics.umd.edu}
\\[0.06in]
{\it Center for String and Particle Theory\\Department of Physics\\ 
University of Maryland \\ 
College Park, MD 20742-4111  USA}\\[.1in]
and  \\ [.1in] 
V.G.J. Rodgers\footnote{vincent-rodgers@uiowa.edu\,\, }
\\[0.06in]
{\it  Department of Physics and Astronomy, 
University of Iowa\\ 
Iowa City, Iowa~~52242--1479 USA}\\[1.8in]

{\bf ABSTRACT}\\[.01in]
\end{center}
\begin{quotation}
{We compute the short distance expansion of  fields or operators  that live in 
the coadjoint representation of an  infinite dimensional Lie algebra by using 
only  properties of the adjoint representation and its dual.  We explicitly compute 
the short distance  expansion for the duals of the Virasoro algebra, affine Lie 
Algebras and the  geometrically realized $N$-extended supersymmetric 
${\cal  GR}$ Virasoro algebra.} 

${~~~}$ \\[.01in] \newline ${~~~\,}$

${~~~~~~~~~~~~~~~~~~~~}$Expansions, 
Supersymmetry, Supergravity
\endtitle
\section{Introduction}  

~~~~The Virasoro algebra is at the heart of understanding string theory
and low dimensional gravitational theories.  In string theory and
conformal field theories it is often thought of as a derived quantity that
comes from the mode expansion  of an energy-momentum tensor.  For
mathematicians it also has meaning in its own right as the one dimensional
algebra of centrally extended Lie derivatives. Representations of the
Virasoro algebra are used to classify conformal field theories and
also provide important clues as to the nature of string field theories.  
One representation, the coadjoint representation has been the focus 
of investigations for two distinct reasons.  One is that the orbits of the 
coadjoint representation under the action of the Virasoro group have 
a relationship with unitary irreducible representations of the Virasoro 
algebra \cite{Segal:ap,Kirillov,Witten:1987ty,Taylor:zp}.  For the Virasoro 
algebra and also affine Lie algebras in one dimension these orbits can 
then be directly related to two dimensional field theories that are conformal 
field theories.  These are the {\em geometric actions} \cite{Rai:1990js,
Alekseev:1989ce,Delius:1990pt}.  Another reason for studying these 
representations comes when one studies the elements of the coadjoint 
representation and adjoint representation as conjugate variables of a field theory \cite{Lano:1994gx}.  
The adjoint elements in these constructions are the conjugate momenta 
if they generate the isotropy algebra of the coadjoint elements.  Field 
theories constructed in this fashion are called {\em transverse actions} 
\cite{Branson:1997pe,Branson:1998bc} (with respect to the 
coadjoint representation) since the geometric actions constructed on the 
orbits are transverse to these transverse actions.  One example of the 
distinction of these two types of actions for the SU(N) affine Lie algebra 
or what physicists sometimes call an SU(N) Kac-Moody algebra is its 
geometric action, an SU(N) WZNW model,and its transverse action, the 
two dimensional SU(N) Yang-Mills action.  For the Virasoro algebra the 
geometric action is given by the two dimensional Polyakov action for gravity 
\cite{Polyakov:1987zb} and its corresponding transverse action given by 
the $N=0$ {\em affirmative  action} \cite{Gates:2001uu,Boveia:2002gf}.   Besides these constructions, other 
areas of interests for the coadjoint representation are the BTZ black holes 
\cite{Banados:2002ey,Banados:1992gq} that appear in the asymptotic 
Brown-Henneaux symmetry \cite{Brown:nw}  on AdS$^3$ 
\cite{Yokoi:1999yv,Nakatsu:1999wt,Banados:2002ey}. 

In this note we will take this primordial view of the algebra and its dual by 
describing a classical phase space where the coadjoint representation 
provides coordinates of a phase space and the adjoint representation are the
conjugate momentum variables.  We will construct Noether charges on 
the phase space which will correspond to the adjoint action on the variables.  
From there it is straightforward to write the short distance expansion for
the elements of the dual representation with themselves or with elements 
of the adjoint representation for that matter.   The algebras of interests to 
us in this note will be the pure Virasoro algebras, the semi-direct product 
of the Virasoro algebra with an affine Lie Algebra, and supersymmetric 
extensions of this for an arbitrary number of supersymmetries.  At this point 
a small review is in order

\section{The Role of the Coadjoint Representation}

~~~~In any space-time dimension the Lie algebra of coordinate transformations 
can be written as 
\begin{equation}
{\cal L}_{\xi} \eta^a = -\xi^b \partial_b \eta^a + \eta^b
\partial_b\xi^a = (\xi \circ \eta)^a,
\end{equation}
and the algebra of Lie derivatives satisfies,
\begin{equation}
 [{\cal L}_{\xi}, {\cal L}_{\eta}] = {\cal L}_{\xi \circ \eta}. 
\end{equation}
In {\em one} dimension, we can centrally extend this
algebra by including  a two cocycle which is coordinate 
invariant and satisfies the Jacobi identity.  We write
\begin{equation}
 [({\cal L}_{\xi}, a), ({\cal L}_{\eta},b) ] = ({\cal L}_{\xi \circ \eta}, (\xi,\eta))
\end{equation}
where the two cocycle depends on the D-dimensional metric $g_{a b}$
(used to define the connection) and a rank two tensor\footnote{While it is possible
to choose $D_{a b}$ = $g_{a b}$, at this stage we do not impose this condition.}
 $D_{a b}$,
\begin{equation}(\xi,\eta) = \frac{c}{2\pi} \int (\xi^a \nabla_a \nabla_b 
\nabla_c\eta^c)\,dx^b + \frac{h}{2\pi} \int (\xi^a D_{a b} \nabla_c \eta^c )\, dx^b -
(\xi\leftrightarrow \eta).
\end{equation}
Here the index structure is left in tact in order to show the
invariance of the two cocycle.
This expression can be viewed as residing on a 0-brane.  
Since  this is a one dimensional structure, 
one may ignore the indices as long as one is mindful of the tensor structure and 
write,
\begin{equation}
(\xi,\eta) = \frac{c}{2\pi} \int (\xi \eta''' - \xi'''
  \eta)\,dx
+ \frac{h}{2\pi} \int (\xi \eta' - \xi' \eta )(D + \frac ch ( \G' + \frac 1{2} \G^2))\,dx,
\label{eq:4}
\end{equation}
where ${}^{\prime}$ is the derivative with respect to the coordinate along the 
0-brane.  In this form, it is easy to see that (\ref{eq:4}) contains the familiar 
``anomaly'' term that arises in string theory.   The two cocycle then can be 
reduced to 
\begin{equation}(\xi,\eta) = \frac{c}{2\pi} \int (\xi \eta''' - \xi'''
  \eta)\,dx
+ \frac{h}{2\pi} \int (\xi \eta' - \xi' \eta )\,B \,\,dx,
\end{equation}
dependent upon a pseudo-tensor, $B=(h D + c\,(\G' + \frac 12 \G^2))$, which 
transforms as 
\begin{equation}
\d B = -2 \xi' B - \xi B' -c\, \xi''' 
\end{equation}
under infinitesimal coordinate transformations.  This pseudo tensor absorbs
the metric contribution in the central extension as well as the tensor
$D$.  $B$ is said to transform in the coadjoint representation of the
Virasoro algebra.  Different choices of $B$ will give different
centrally extended algebras.  In string theory, central extensions are
commonly chosen so that an SL(2,R) subalgebra is centerless.  
Thus the coadjoint representation might be thought of as endemic to
the central extension of the algebra. For example, $B=0$, is the
choice commonly used when the metric is  fixed to $g_{a b} = 1$ and
$D_{a b} = 0$.  The 
vector fields $\xi$ for the one dimensional line are moded by $\xi =
\sum C_N x^{N+1}$.  Then the realization for the Virasoro algebra is 
\begin{equation}
\left[ L_N,L_M\right]  =(N-M)\,L_{N+M} 
+ (c N^3 - c  N)\,\delta _{N+M,0}.
\end{equation}
One can get the same central extension for the algebra on the circle in two 
distinct ways. Either by making a change of coordinates where $x = \exp{(i \o)}$ 
giving the complex metric $g(\o) = \exp{(2 i \o)}$ or by choosing $g(\t) = 1$ and 
$D_{a b} = g_{a b}$ on the circle with $h=-c$.  In this case the vector fields are 
$\xi = \sum\,C_N \,\exp{( i N \o)}$. 

As a parenthetic  remark, one may wish to view the two cocycle $(\xi,\eta)$ as 
a functional of the metric.  Upon variation of $(\xi,\eta)$ with respect to the metric
$g_{a b}$ and assuming that $D_{a b}$ is independent of the metric, one finds
\begin{equation}
{\d \over \d g} (\xi, \eta) = {\rm J}'\,\G + {\rm J}\, \G',
\end{equation}
where the current is, ${\rm J} = \eta \,\xi' - \eta'\, \xi$.
For constant ${\rm J}$, one recognizes this variation as the anomalous one
dimensional ``energy-momentum'' tensor.  Examples of higher dimensional
versions of non-central extensions may be found in \cite{Larsson:vk,Larsson:fx,Dzhumadildaev:sf}.

Another way in which the coadjoint representation appears is directly
through the construction of a representation that is dual to the adjoint 
representation.  In this case one starts with the centrally extended algebra.  
Then using a suitable pairing between the algebra and its dual, one 
extracts the coadjoint action of the algebra.  As an example consider 
the semi-direct product of the Virasoro algebra with an affine Lie algebra.  
Then the Virasoro algebra
\begin{equation}
\left[ L_N,L_M\right]  =(N-M)\,L_{N+M} 
+ c N^3\,\delta _{N+M,0} 
\end{equation}
is augmented with 
\begin{equation}
\left[ J_N^\alpha ,J_M^\beta \right]  =i\,f^{\alpha \beta \gamma
}\,J_{N+M}^\beta +N\,k\;\delta _{N+M,0}\;\delta ^{\alpha \beta }
\end{equation}
and 
\begin{equation}
\left[ L_N,J_M^\alpha \right]  = -M\;J_{N+M}^\alpha. 
\end{equation}
where 
$ [\tau^\alpha, \tau^\beta] = i f^{\alpha \beta \gamma} \tau^\gamma. $
The algebra is realized by 
\begin{equation}
L_N= \xi_{N}^a \partial_a = ie^{iN\theta }\partial _\theta, \,\,\,\,\,\,\,J_N^\alpha 
= \tau ^\alpha e^{iN\theta },
\end{equation}
so that the centrally extended basis can be thought of as the three-tuple,
\begin{equation}
\left(L_A,J_B^\beta ,\rho\right).
\end{equation}
The adjoint representation acts on itself as
\begin{equation}
\left( {L_A,J_B^\beta ,\rho }\right) {*}\left( {L_{N^{\prime }},J_{M^{\prime
}}^{\alpha ^{\prime }},\mu }\right)  =\left( {L_{new},J_{new},\lambda }%
\right) \label{case1-a}
\end{equation}
where
\begin{eqnarray}
L_{new}&=&  \,(A-N^{\prime })\,L_{A+N^{\prime }} \cr
J_{new}&=&  -M^{\prime } J_{A+M^{\prime }}^{\alpha ^{\prime
}}+BJ_{B+N^{\prime }}^{\,\beta }+if\,^{\beta \alpha ^{\prime }\lambda
}J_{B+M^{\prime }}^\lambda\cr
\lambda &=&(c A^3) \delta _{A+N^{\prime },0}+Bk\delta ^{\alpha
^{\prime }\beta }\delta _{B+M^{\prime },0}. \label{case1-b}
\end{eqnarray}

A typical basis for the dual of the algebra can be written as the
three-tuple
  $\left( \widetilde{L_N},\widetilde{J_M^\alpha },\widetilde{\mu }\right)$ .

Using the pairing,
\begin{equation}
\left\langle \left( \widetilde{L_N},\widetilde{J_M^\alpha },\widetilde{\mu }%
\right) \left| \left( L_A,J_B^\beta ,\rho \right) \right. \right\rangle
=\delta _{N,A}+\delta ^{\alpha \beta }\delta _{M,B}+\rho \widetilde{\mu }
\end{equation} 
and requiring it to be invariant
one defines  the  coadjoint representation through the action of the
adjoint on this dual as \cite{Lano:tc}
\begin{eqnarray}
\left( {L_A,J_B^\beta ,\rho }\right) &{*}&\left( {{\tilde L_N},{\tilde
J_M^\alpha },{\tilde \mu }}\right) =\left( {{\tilde L_{new}},{\tilde
J_M^\alpha },0}\right) \mbox{ } \,\,\,\,\,\,\,\mbox{\rm with,}\\ \cr
{{\tilde L_{new}}}{}{=}&& {(2A-N){\tilde L_{N-A}} {-} B\delta ^{\alpha \beta }{%
\tilde L_{M-B}}} -{\tilde \mu}(c A^3){\tilde
L_{-A}}\,\,\,\,\,\,\,\,\, \mbox{\rm and} \\\cr
{\tilde J_M^\alpha }{}{=} &&{(M-A){\tilde J_{M-A}^\alpha }-if\,^{\beta \nu
\alpha }{\tilde J_{M-B}^\nu }} -{\tilde \mu} B\,k\,{\tilde J_{-B}^{\,\beta }.%
} 
\label{case1-c}
\end{eqnarray}
Since we are interested in field theories it is more instructive to use explicit tensors
instead of the mode decomposition.  The adjoint representation may be
though of as three-tuple, 
\begin{equation}{\cal F}=\left( \xi \left( \theta \right), \Lambda \left(
\theta \right) ,a\right)\label{eqn1}
\end{equation}
 containing a  vector field $\x^a$, coming from the Virasoro algebra, a
gauge parameter $\L$ coming from the affine Lie algebra and a central
extension $a$.  The coadjoint element is the three-tuple,
\begin{equation} {\rm B} =\left( {\rm D}\left( \theta \right), {\rm A}\left(
\theta \right), \mu \right),\label{eqn2} 
\end{equation}
 which consists of a rank two
pseudo tensor $D_{a b}$, a gauge field $A_a$ and a corresponding
central element $\mu$.  In this way the coadjoint action can be
written as 

\begin{eqnarray}
\delta \widetilde{{\rm B}_F}&&=\left( \xi \left( \theta \right), \Lambda \left(
\theta \right) ,a\right) *\left( {\rm D}\left( \theta \right), {\rm A}\left(
\theta \right), \mu \right) \cr
&& =\left( \delta {\rm D}\left( \theta \right), \delta {\rm A}\left( \theta 
\right), 0\right) \mbox{,} 
\end{eqnarray}
\begin{equation}
\delta {\rm D}\left( \theta \right) =\;
\stackunder{\rm coordinate\ transformation}{\underbrace{2\xi ^{^{\prime
}}{\rm D}+{\rm D}^{^{\prime }}\xi + \frac {c \mu}{2\pi}\xi^{\prime \prime
\prime } + \frac {h \mu}{2 \pi} \xi^{\prime}}}-\stackunder{\rm gauge\ trans}
{\underbrace{Tr\left( {\rm A}\Lambda^{\prime }\right)}} \label{variation_D}
\end{equation}
and
\begin{equation}
\delta {\rm A(\theta )}=\;
\stackunder{\rm coord\ trans}{\underbrace{{\rm A}^{\prime }\xi +\xi^{\prime }
{\rm A}}}-\stackunder{\rm gauge\ transformation}{\,\underbrace{[\Lambda
\,{\rm A}-{\rm A\,}\Lambda ]+k\,\mu \,\Lambda^{\prime
}}}. \label{variation_A}
\end{equation}
Again  $^{\prime }$ denotes derivative with respect to the argument.

\section{As Phase Space Variables}

~~~~The role of the adjoint and coadjoint representations as phase space
elements is best represented in a familiar example with emphasis on the 
algebraic structure.  Consider the phase space elements of a Yang-Mills 
theory.  There one has the  vector potential as a canonical coordinate,
$A_i^a(x)$ and the electric field $E_i^a(x)$ as its  conjugate momentum 
through the Poisson bracket relations
\begin{eqnarray}
\lbrack A_i^a(x),A_j^b(y)] &=&0  \cr
{\lbrack E_i^a(x),E_j^b(y)]} &=&0 \cr
{\lbrack A_i^a(x),E_j^b(y)]} &=&i\delta ^{ab}\, \d_{ij}\,\delta (x,y).
\end{eqnarray}
The transformation laws for these phase space variables under spatially 
dependent gauge transformations is given by
\begin{eqnarray}
A_i(x) &\rightarrow &U(x)A_iU^{-1}(x)-{\frac 1g}\partial _iU(x)U^{-1}(x)
\cr
E_i(x) &\rightarrow &U(x)E_iU^{-1}(x), 
\end{eqnarray}
where one recognizes that $A_i(x)$ (when reduced on a 0-brane) is the 
second element of the three-tuple   in Eq(\ref{eqn2}) of the  element of 
the coadjoint representation of an SU(N) affine Lie algebra on the line 
and $E_i(x)$ is in the adjoint representation (the second element in 
Eq(\ref{eqn1}). In this way one sees that these representations which 
are dual to each other are indeed phase space elements.  

If we were to consider the  generating function for spatial gauge transformations, 
\begin{equation}
G(x)^a=\partial _iE_i^a+[E_i,A_i]^a,
\end{equation} 
then one can construct the  charge
\begin{eqnarray}
Q_{\L}&=&\int dx\ G^a\Lambda ^a\left( x\right)
\,\,\,\,\,\,\,\,\,{\rm with}\\
\cr
\{Q_{\L},E(x)\} &=&[\Lambda(x) ,E(x)]  \cr
\{Q_{\L},A(x)\} &=&[\Lambda(x) ,A(x)]-\frac 1g\partial \Lambda(x)
\end{eqnarray}
For a generic function on the phase space, say $F(A,E)$
\begin{equation}
\{Q_{\L}, F(A,E) \} = \L(x) {\d \over \d {\hat \L(x)}}\, F(A,E) \label{eqn3}
\end{equation}
where ${\d \over \d {\hat \L(x)}}$ is the functional variation in the
direction of $\L(x)$.

Similarly, the Virasoro algebra and its dual produce a set of phase space variables.  
Let the one dimensional pseudo tensor, $D_{i j}$ (when reduced on a 0-brane) correspond 
to the first element in Eq.(\ref{eqn2}) with conjugate momentum given by the rank 
two tensor density of weight one, $X^{i j}$.   The generator corresponding 
to the one dimensional coordinate transformation is given by 
\begin{equation}
{G}_a(x) = X^{l m}\, \partial_a D_{l m} 
-\partial_l(X^{l m}\, D_{a m}) -\partial_m (D_{l a}\, X^{l m}) 
- c~ \partial_a \partial_l \partial_m X^{l m}. \label{tict}
\end{equation}
>From the Poisson brackets, one can recover the transformation 
laws of $X^{a b}$ and $D_{a b}$.  We have
\begin{equation}
Q_{\xi}=\int dx\ G_a \xi^a,  
\end{equation}
where $\xi^a$ (when reduced on a 0-brane) corresponds to $\xi$ in the algebra.  One has
\begin{eqnarray}
\{Q_{\xi},\,D_{l m}(x) \}&=& -\xi^a(x) \,\partial_a 
D_{l m}(x) - D_{a m}(x) \,\partial_l \xi^a(x)\,-\,D_{l a}(x)\, \partial_m \xi^a(x)
 - c~\,\partial_a \partial_l \partial_m
\,\xi ^a(x) \nonumber  \\
&=& -2 \xi'(x) \,D(x) - \xi(x) \,D'(x) - c\, \xi(x)''' \cr
\{Q_{\xi},\,X^{l m}(x)\} &=&\xi^a(x) \partial_a \,X^{ l m}(x) - 
(\partial_a \xi^l(x)) \,X^{a m}(x) - (\partial_a \xi^m(x) ) \,X^{l a}(x) + 
\,(\partial_a \xi^a(x)) X^{l m}(x)\nonumber \\
&=& \xi(x)\, X'(x) -\xi'(x)\, X(x)  \label{gaussdiff}
\end{eqnarray}
Again for a generic function on the phase space, say $F(D,X)$
\begin{equation}
\{Q_{\xi}, F(D,X) \} = \xi(x)\, {\d \over \d {\hat \xi(x)}}\, F(D,X) \label{eqn4}
\end{equation}
where ${\d \over \d {\hat \xi(x)}}$ is the functional variation in the
direction of $\xi(x)$.

\section{Short Distance Expansions from Algebras}

~~~~It is now straightforward to extract the short distance expansion
\cite{Wilson} for
elements of the coadjoint representation.  Equations (\ref{eqn3}) and
(\ref{eqn4}) are the prototype expressions that we need to proceed.
For any element of the adjoint representation, say $\a$, we can
construct $Q_\a$.  Now $\a$ is dual to coadjoint elements, say ${\cal A}$ since 
\begin{equation}
(\a, {\cal A}) = \int \a(x) {\cal A}(x) \, dx =  \,\,\,\,\,\,{\rm some\ constant}.
\end{equation}
We treat the above expression as the integral form of the
operation with $\d \over \d {\hat \a}$.  Then  for any function of the phase space $F(x)$
\begin{eqnarray}
\{Q_{\a}, F(x) \} &=& \a(x)\, {\d \over \d {\hat \a(x)}}\, F(x) \\ \cr
&=& \int \a(y) \left({\cal A}(y) G(x) \right) \,dy. \label{eqn5}
\end{eqnarray}
The distribution $f_{A}(x,y)={\cal A}(y) F(x) $ when compared to the
left hand side of Eq(\ref{eqn5}) gives the short distance expansion
between ${\cal A}(x)$ and $F(x)$.  
In 
the following we derive short distance expansions for some simple cases 
and then move on to the N-extended super Virasoro algebra.

\subsection{Case:  Virasoro and Affine Lie Algebras}
\label{case1}

~~~~As our first example we consider the short distance expansion for the semi
direct product of the Virasoro algebra  and an affine Lie algebra.
We are interested in the short distance expansion of the coadjoint
elements $D(x)$ and $A(x)$.  Equations \ref{variation_D} and
\ref{variation_A} allow us to construct the charge 
\begin{equation}
Q_{\xi} = \int \, X(y) \left(2 \xi'(y)\, D(y) + D(y)'\, \xi + {i c \mu
  \over 12} \, \xi'''\right) + \int\, E(y)\left( \xi'(y) A(y) + A'(y)
\xi(y) \right).
\end{equation}

Then from $\{ Q_{\xi}, \, A(x)\}$ we have that 
\begin{equation}
\{ Q_{\xi}, \, A^b(x)\} = \xi(x) {\d \over \d {\hat \xi(x)}} A^b(x) = \int \,
\xi(y) \, D(y)\, A^b(x) \, dy\,\,.
\end{equation}
This implies that 
\begin{equation}
D(y)\, A^b(x) = (\partial_y \d(x,y))\, A^b(x) - (\partial_x A^b(x)) \,
\d(x,y).
\end{equation}
Using the representation for the delta function on a line 
\begin{equation}
\d(x,y) = {1 \over2 \pi i (y-x)},
\end{equation}
we have that 
\begin{equation}
D(y)\, A^b(x) = {-1 \over 2 \pi i (y-x)^2} \, A^b(x) - {1\over2 \pi i (y-x)}
\partial_x A^b(x).
\end{equation}

Similarly we can construct a charge $Q_\L$ via 
\begin{equation}
Q_\L = \int \left( - X(y) A^b(y) {\L'^{b}}(y) - i f^{b a c}
  E^b(y)\L^a(y) A^c(y) + k \mu \,  E^b(y)\,{\L'^b}(y) \right) \,dy
\end{equation}
then together $Q_L$ and $Q_\xi$ will give 
\begin{eqnarray}
D(y)\, D(x) &=& {-1 \over 2 \pi i (y-x)} \partial_x D(x) + {1 \over 
  \pi i (y-x)^2} D(x) - {c \over 2 \pi i (y-x)^4} \\ \cr
D(y) \, A^b(x) &=& {-1 \over 2 \pi i (y-x)^2} A^b(x) - {1 \over 2 \pi i
  (y-x)} \partial_x A^b(x) \\ \cr
A^b(y) \, D(x) &=& {1 \over (y-x)^2} \, A^b(x) \\ \cr
A^b(y) \, A^a(x) &=& {i k \mu \over (y-x)^2} \d^{b a} - i\, f^{b a c}
{1 \over (y-x)} A^c(x).
\end{eqnarray}

\subsection{Case:  $N$=1 Super Virasoro Algebras}

~~~~Supersymmetric algebras can be treated in a similar way.  From the
discussion  found in Siegel \cite{Siegel:1999ew}, let
$\Phi^M$ represent an element of the phase space and let $\O^{M N}$ be
a supersymplectic two form with an inverse $\O_{M N}$, i.e. 
\[\O^{M N} \,  \O_{P N} = \d^M_P.\]  The index $M = \{m, \m \}$ 
where the Latin indices are bosonic and the Greek indices fermionic.  
Then 
\begin{equation}
[ \Phi^M, \Phi^N \} = \, \hbar \, \O^{M N},
\end{equation}
and 
\[ \O_{\{M N]} = 0.
\]
This means that \[ \O_{(m n)} = \O_{[\m \n]}=\O_{m\n} + \O_{\n m} =
0. \] The bracket is defined by  
\begin{equation}
[ A, B \} = - A \,{\stackrel{\leftarrow}{\partial}\over \partial
  \Phi^M} \, \O^{N M} \, {{\partial B}\over \partial \Phi^N}
\end{equation}

For the $N$=1 super Virasoro algebra, the adjoint representation is built 
from the vector field $\xi$, a spinor field $\e$ and and a central extension 
so that we might write an adjoint element as ${\cal F} = (\xi, \e, a)$.  In the 
same way the coadjoint representation is given by a three-tuple of fields, 
$B=(D, \psi, \a)$.  The transformation law for the coadjoint representation 
is 
\begin{equation}
{\cal F}*B= \left(-\xi\, D' -\frac 12 \e\, \psi' - \frac 32 \e'
\psi - \frac {c\, \b}8\, \xi''', \, - \xi \, \psi' - \frac 12 \e\,\psi\,  -\frac
32 \xi' \, \psi - 4i \b c \, \e'', \,0\right).
\end{equation}
The charges are 
\begin{eqnarray}
Q_\xi &=& \int \, X(y)\left( -\xi(y) \,D(y) \, - \frac {c\, \b}8\, \xi'''
\right)\,dy + \int \, \phi(y)\left(  - \xi \, \psi(y)' -\frac32 \xi' \, 
\psi(y)\right)\,dy,\\
\,{\rm and} \cr
Q_\e &=& \int \, X(y)\left( -\frac 12 \e(x) u(y)' - \frac 32
  \e'(y)\,\psi(y)\right)\, dy + \int \, \phi(y)\left( - \frac 12 \e(y)
  - 4i \b c \,
  \e'' \right) \, dy
\end{eqnarray}
Here the fields $X(y)$ and $\phi(y)$ correspond respectively to the spin
$-1$ and spin $-\frac 12$ conjugate momenta for $D$ and $\psi$,
which gives the short distance expansion
\begin{eqnarray}
D(y)\, D(x) &=& {-1 \over 2 \pi i (y-x)} \partial_x D(x) + {1 \over 
  \pi i (y-x)^2} D(x) - {3 c \b \over 8 \pi i (y-x)^4} \\ \cr
D(y) \, \psi(x) &=& {-3\over 4 \pi i (y-x)^2} \psi(x) - {1 \over 2 \pi i
  (y-x)} \partial_x \psi(x) \\ \cr
\psi(y) \, D(x) &=& {3 \over 4 \pi i (y-x)^2} \, \psi(x) - {1 \over 4
  \pi i (y-x)} \psi(x) \\ \cr
\psi(y) \, \psi(x) &=& {-2  \over \pi (y-x) } D(x)  - {4 \b c \over
  \pi (y-x)^3}.
\end{eqnarray}

\section{$N$-Extended ${\cal GR}$ Super Virasoro Algebra} 

~~~~We now come to the final example and the main result of this work, the
$N$-extended Super Virasoro algebras.  The $N$ Extended super Virsoro
algebra that we will use is the one found ing \cite{Gates:1995ch,Gates:1998ss}.  
The generators of this algebra provide an (almost\cite{Curto:2000kd}) primary basis to
the $K(1\mid N)$ contact superalgebra \footnote{We thank Thomas
  Larsson for pointing this out to us.}.  This is the subalgebra
of ${\rm vect}(1\mid N)$ vector fields that preserves the contact one form 
\[ \s = d\t + \d_{I J} \z^I d\z^J. \] The generators are 
\[ {\cal Q}^{\rm I_1 \cdots I_p }_{\cal A} \, =\, \t^{\cal A}
\z^{\rm I_1} \cdots \z^{\rm I_p} \partial_\t ~~~,~~~
{\cal P}^{\rm I_1 \cdots  I_{p+1}}_{\cal A} \, = \, \t^{\cal A}
\z^{\rm I_1} \cdots \z^{\rm I_p} \partial ^{\rm I_{p+1}} 
~~~,\]   where ${\cal A}$ can be integer or
half-integer moded. This algebra along with a
complete classification of Lie superalgebras used in string theory are
reviewed and discussed in \cite{GLS97}.  
In the ``almost'' primary basis the generators may be written as \footnote{We are grateful to Thomas
  Larsson for pointing out errors in the algebra in previous publications}: 
the following:
\begin{eqnarray}
G_{\cal A} {}^{\rm I} &\equiv& i \, \t^{{\cal A} + \fracm 12}  \,
\Big[\, 
\, \pa^{\rm I} ~-~ i \, 2 \, \zeta^{\rm I} \, \pa_{\t} ~ \Big] ~+~ 2 (\, 
{\cal A} \,+\, \fracm 12 \,) \t^{{\cal A} - \fracm 12}  \z^{\rm I}
\z^{\rm 
K} \, \pa_{\rm K} ~~~, \\
L_{\cal A} &\equiv& -\, \Big[ \, \t^{{\cal A} + 1} \pa_{\t} ~+~ \fracm
12
({\cal A} \, + \, 1) \, \t^{\cal A} \zeta^{\rm I} \, \pa_{\rm I} ~
\Big]  
~~~, \\
U_{\cal A}^{{\rm I}_1 \cdots {\rm I}_q} &\equiv& i \, (i)^{ [\fracm q2
]} 
\, \t^{({\cal A} - \fracm {(q - 2)}2 )} \zeta^{{\rm 
I}_1} \cdots \, \zeta^{{\rm I}_{q-1}} \, \pa^{{\rm I}_q} ~~~,~~~ q \, =
\, 
1 , \, \dots \, , \, N \, + \, 1 ~~~,  \\
{\cal R}{}_{\cal A}^{{\rm I}_1 \cdots {\rm I}_p} &\equiv&  (i)^{ [\fracm
p2 ]} \, \t^{({\cal A} - \fracm {p}2 )} \zeta^{{\rm I}_1} \cdots \, 
\zeta^{{\rm I}_p}  \, \t \pa_{\t}  ~~~~,~~~ p \, = \, 2 , \, \dots \, , \, N ~~~
\end{eqnarray}
for any number $N$ of supersymmetries. The antisymmetric part of the 
$U_{\cal A}^{{\rm [I} \, {\rm J]}}$ generator will generate an SO(N)
subalgebra so we will denote it as 
\[
T_{\cal A}^{\, \rm {I \, J }} \equiv  - U_{\cal A}^{{\rm [I} \, {\rm
    J]}}. \]  
Note that the primary genrators 
 $L_{\cal A}$ and $G_{\cal A} {}^{\rm I}$ are used in lieu of 
${\cal R}{}_{\cal A}$ and ${\cal R}{}_{\cal A}^{{\rm I}_1}$
respectively while ${\cal R}{}_{\cal A}^{{\rm I}_1 {\rm I}_2}$ is
the only generator that is not primary. 

The elements of the algebra can be realized as fields whose tensor
properties can be determined by they way they transform under 
one dimensional coordinate transformations. How each field transforms
under a Lie derivative with respect to $\xi$  is summarized below. In
conformal field theory these transformation laws are generalized for
two copies of the diffeomorphism algebra and characterize the
fields by weight and spin.  Here we treat the algebraic elements as one
dimensional tensors.

\begin{center}
\begin{tabular}{|c|c|c|}\hline
\multicolumn{3}{|c|}{ \bf Table 1: Tensors Associated with the Algebra} 
\\ \hline\hline
${\rm {Element\, of\, algebra}}$ & ${\rm {Transformation\, Rule}}$ & 
${\rm Tensor\, Structure}$ \\ \hline 
$ L_{\cal A} ~\to~ \eta $ &  $ \eta  ~\to~ -\xi' \eta + \xi \eta'  
~~~~$ & $ \eta^a$  \\ \hline
$ G^{\rm I}_{\cal A} ~\to~ \chi^{\rm I} $ &  $ \chi^{\rm
  I}~\to~{-\xi(\chi^{\rm I})' +\fracm 12 \xi' \chi^{\rm I}} ~~ $
&$\chi^{\rm I; \a}$  \\ \hline
$T^{\rm{ R} { S}}~\to~t^{\rm{ R}{ S}} $ &  $ t^{\rm{ R}{ S}}  ~\to~  
-\xi\,({t}^{\rm{ R}{S}})'~~~~$ & $ t^{\rm{ R}{ S}} $  \\ \hline
$U^{\rm V_1 \cdots V_n} ~\to~ w^{\rm V_1 \cdots V_n}  $ &  $ w^{\rm
  V_1  \cdots  V_n}\rightarrow  -\xi (w^{\rm V_1
  \cdots  V_n})'  -\fracm 12 (n-2) \xi' w^{\rm V_1 \cdots V_n}~~~~$ &
$w^{\rm V_1 \cdots  V_n;\, a}_{\,\,\,\,\,\,\,\,\,\,\,\, \a_1 \cdots \a_n}  
$  \\ \hline
$R^{\rm T_1 \cdots T_n}_{\cal A}~\to~ r^{\rm T_1 \cdots T_n} $ &
${r}^{\rm T_1 \cdots T_n} ~\to~   -({r}^{\rm T_1 \cdots T_n})'\xi-
\fracm 12 (n-2)  \xi'  r^{\rm T_1 \cdots T_n}  ~~~~$ & $ r^{{\rm
    T_1 \cdots T_n}\, ;a }_{\,\,\,\,\,\,\,\,\,\,\,\,  \a_1 \cdots \a_n} $ 
\\ \hline
\end{tabular}
\end{center}
In the above table we have used capital Latin letters, such as ${\rm I,J,K}$
to represent SO($N$) indices, small Latin letters to represent tensor indices, 
and small Greek letters for spinor indices.   Spinors with their indices up 
transform as scalar tensor densities of weight one (1) while those with their 
indices down transform as scalar densities of weight minus one (-1).  For 
example, the generator $U^{\rm V_1 V_2 V_3}$ has a tensor density realization 
of contravariant tensor with rank one and weight $-\frac{3}{2}$ living in 
the ${\rm N \times  N \times N}$ representation of SO($N$), i.e. $\o^{{\rm 
V_1 V_2 V_3}; a}_{\a_1 \a_2 \a_3}$. 

\begin{center}
\begin{tabular}{|c|c|c|}\hline
\multicolumn{3}{|c|}{ \bf Table 2: Tensors Associated with the Dual of the  
Algebra} \\ \hline\hline
${\rm {Dual \,element\, of\, algebra}}$ & ${\rm {Transformation\, Rule}}$ & 
${\rm Tensor\, Structure}$ \\ \hline 
$ {L^\star}_{\cal A} ~\to~ D $ &  $ D  ~\to~ -2 \xi' D - \xi D'  ~~~~$
& $ D_{a b}$  \\ \hline
$ {G^\star}^{\rm I}_{\cal A} ~\to~ \psi^{\rm I} $ &  $ \psi^{\rm
  I}~\to~{- \xi(\psi^{\rm I})' - \fracm 32 \xi' \psi^{\rm I}} ~~ $
&$\psi^{\rm I}_{a \a}$  \\ \hline
${T^\star}^{\rm{ R} { S}}~\to~A^{\rm{ R}{ S}} $ &  $ A^{\rm{ R}{ S}}
~\to~  -(\xi)' A^{\rm{ R}{ S}} -\xi\,({A}^{\rm{ R}{S}})'~~~~$ & $ A^{\rm{ 
R}{ S}}_a $  \\ \hline
${U^\star}^{\rm V_1 \cdots V_n} ~\to~ \o^{\rm V_1 \cdots V_n}  $ &  $
\o^{\rm V_1 \cdots V_n}~\to~  -\xi (\o^{\rm V_1  \cdots  V_n})'  -(2
-\fracm n2) \xi' \o^{\rm V_1 \cdots V_n}~~~~$ &
${\o}^{\rm V_1 \cdots  V_n;\, \a_1 \cdots \a_n}_{a b}  $  \\ \hline
${R^\star}^{\rm T_1 \cdots T_r}_{\cal A}~\to~ \rho^{\rm T_1 \cdots T_r} $ &
${\rho}^{\rm T_1 \cdots T_r} ~\to~   -({\rho}^{\rm T_1 \cdots T_r})'\xi-
(2-\frac r2)  \xi'  \rho^{\rm T_1 \cdots T_r}  ~~~~$ & $ \rho^{{\rm
    T_1 \cdots T_r}; \a_1 \cdots \a_r}_{a b} $ \\ \hline
\end{tabular}
\end{center}

Thus for $N$ supersymmetries there is one rank two tensor $D_{a b}$,
$N$ spin-$\fracm 32$ fields $\psi^{\rm  I},$ a spin-1 covariant tensor 
$A^{\rm {R}{S}}$ that serves as the $N (N - 1)/2$ SO($N$) vector 
potentials (that are gauge potentials for $N=2$) associated with the
supersymmetries, $N \, (\, 2^N)$ fields for the $ {\o^{\rm {V_1} \cdots {V_p}}}$
and $N \, (\, 2^N - \,N - \,1)$ $ {\r^{\rm {T_1} \cdots {T_p}}}$ fields.  The
entries in the third column of each of these tables are the 0-brane
reduced tensors or tensor densities corresponding to the ones that
appear in the second column.
\subsection{Central Extensons of the $N$-Extended {\cal GR} algebra}

In \cite{Curto:2000kd,Boveia:2002gf} it was mistakenly reported that
the $N$-Extended {\cal GR} algebra admits a central extension for
arbitrary values of $N$ and further the $U^{I_1}_{\cal A}$ and the symmetric
combination for $U^{(I_1 I_2)}_{\cal A}$ generators where omitted.  It is
easy to see that not only are these required to close the algebra but
their existence will eliminate the central extensions for $N >
2$.   Although the algebra is not simple since the $U^I_{\cal A}$
fields form an abelian extension to the algebra the obstruction comes
from the required symmetic part of $U^{J_1 J_2}_{\cal A}$.  This can be
seen by looking at the $(G^I_{\cal A}, U^{I J}_{\cal B}, G^K_{\cal C})$
contribution to the Jacobi identity.  There one finds that 
\[ [  U^{J_1 J_2}_{\cal B}, \{G^I_{\cal A}, G^K_{\cal C}\}] - \{
G^K_{\cal C}, [ U^{J_1 J_2}_{\cal B}, G^I_{\cal A}] \} + \{G^I_{\cal A},
[ G^K_{\cal C}, U^{J_1 J_2}_{\cal B}] \} =0 \]
implies that 
\[ -2 i(A-C)\,[ U^{J_1 J_2}_{\cal B}, \,U^{I K}_{\cal A + C} ] - i
{\tilde  c} \,\big((C^2 - \frac 14) \d^{J_1 K} \d^{I J_2} -
(A^2 - \frac 14) \d^{ J_2 K} \d^{I j_1}\big) \d_{A+B+C} = 0 .\]
It is clear that the symmtric combination cannot satisfy the Jacobi
identity.  As we will show below, only the antisymmtric fields
$U^{[I_1 I_2]}_{\cal A}$ are needed to close the algbra and a central
extension exists.  The constraint above forces the central extension
in the SO(2) affine Kac-Moody algebra to be related to the central
extension in the superdiffeomorphism algebra even before unitarity
issues are considered.   In what follows we will
seperate the $N=2$ algebra from the generic case.

\subsection{$N=2$}
For $N=2$ the ${\cal GR}$ Super Virasoro Algebra
reduces to a smaller set of generators that close as well as 
admits a central extension.  The algebra is 
\begin{eqnarray}
[~ L_{\cal A} \, , \, L_{\cal B} ~\} \,  &=& ( \, {\cal A} \, 
- \, {\cal B} \, ) \, L_{{\cal A} + {\cal B}} + \fracm {1}{8} 
\,c \,({\cal A}^3-{\cal A})\, \d_{{\cal A} + {\cal B},0}~~~,\\ \cr
[ ~ G_{\cal A}{}^{\rm I} \, , \, G_{\cal B}{}^{\rm J} ~\} \, &=& 
-\, i \, 4 \,  \d^{{\rm {I\, J}}} L_{{\cal A} + {\cal B}}
~-~ i 2 ({\cal A} - {\cal B} ) \, [ \, \, T_{{\cal A} + {\cal B}
}^{\rm {I\, J}} ~+~ 2  ({\cal A} + {\cal B} ) \, U_{{\cal A} + 
{\cal B}}^{\rm {I\, J \,K}}{}_{\rm K} ~ ]   \cr
 &&\,\,\,\,\,-i  c ({\cal A}^2 -\fracm 14 ) \, \d_{{\cal 
A+ B}, 0} \, \d^{\rm I\, J}~~~,\\ \cr
[ ~ L_{\cal A} \, , \, G_{\cal B}{}^{\rm I} ~\} \,  &=& ( \, 
\fracm 12 {\cal A} \, - \, {\cal B} \, ) \, G_{{\cal A} + {\cal 
B}}{}^{\rm I}~~~,\\ \cr
[ L_{\cal A} \, ,\, T_{\cal B}^{\, \rm {I \, J }} \} \, &=&~ - \,{\cal B} 
\, \,T_{\cal A+B}^{\, \rm {I \, J }} ~~~, \\ \cr
[~ T_{\cal A}^{\, \rm {I \, J }} \,,\,  T_{\cal B}^{\, \rm {K \, L }}
~\} 
&=& T_{\cal A+B}^{\, \rm {I \, K }} \, \d^{\rm J L}+~ T_{\cal A+B}^{\, 
\rm {J \, L }} \, \d^{\rm I K}-~ T_{\cal A+B}^{\, \rm {I \, L }} \,
\d^{\rm J K}-~ T_{\cal A+B}^{\, \rm {J \, K }} \, \d^{\rm I L} \cr
&&~~~~~-2 c \, ({\cal A}-{\cal B}) (\d^{\rm I K} \d^{\rm J L}\, 
- \d^{\rm I L}  \d^{\rm J K}) \d_{{\cal A} +{\cal B},0} ~~~,\\ \cr
[~ T_{\cal A}^{\, \rm {I \, J }}\, , \, G_{\cal B}{}^{\rm K} ~\} \, 
&=&~ 2\,(\d^{\rm J K} G_{\cal A+B}{}^{\rm I} \, - \d^{\rm I K} G_{\cal 
A+B}{}^{\rm J})
%%&&~~~~+~ 2 {\cal A} \, (\d^{\rm J K} \, \, U_{{\cal A} + 
%{\cal B}}^{\rm {I \, , L}}{}_{\rm L} - \d^{\rm I K} \, \, U_{{\cal A} 
%+ {\cal B}}^{\rm {J \, L}}{}_{\rm L}  + \, U_{{\cal A} + {\cal B}}^{
%\rm {J \, K \, I\,}} - \, U_{{\cal A} + {\cal B}}^{\rm {I\, K \, J}} 
%) ~~~.
\end{eqnarray}
Where 
\[
T_{\cal A}^{\, \rm {I \, J }} \equiv  - U_{\cal A}^{{\rm [I} \, {\rm
    J]}} \]
serves as the SO(2) generator.  

\subsection{ $N > 2$} 
The $N > 2$ the central extension is absent and new generators must be
included including the ``low order'' generators $ U_{{\cal A}}^{\rm
  I}$.  
The commutation relations are
\begin{eqnarray}
[~ L_{\cal A} \, , \, L_{\cal B} ~\} \,  &=& ( \, {\cal A} \, 
- \, {\cal B} \, ) \, L_{{\cal A} + {\cal B}} \\~~~\cr
[ ~ L_{\cal A} \,, \, U_{\cal B}^{{\rm I}_1 \cdots {\rm I}_m}
~ \} &=& - \, [ ~ {\cal B} \, + \, \fracm 12 \, (m-2) \, {\cal 
A} ~] \, U_{\cal A + \cal B}^{{\rm I}_1 \cdots {\rm I}_m} ~~~,\\
\cr
[ ~ G_{\cal A}{}^{\rm I} \, , \, G_{\cal B}{}^{\rm J} ~\} \, &=& 
-\, i \, 4 \,  \d^{{\rm {I\, J}}} L_{{\cal A} + {\cal B}}
~-~ i 2 ({\cal A} - {\cal B} ) \, [ \, \, T_{{\cal A} + {\cal B}
}^{\rm {I\, J}} ~+~ 2  ({\cal A} + {\cal B} ) \, U_{{\cal A} + 
{\cal B}}^{\rm {I\, J \,K}}{}_{\rm K} ~ ]~~,\\ \cr
[ ~ L_{\cal A} \, , \, G_{\cal B}{}^{\rm I} ~\} \,  &=& ( \, 
\fracm 12 {\cal A} \, - \, {\cal B} \, ) \, G_{{\cal A} + {\cal 
B}}{}^{\rm I}~~~,\\ \cr
[ ~ L_{\cal A} \,, \, R_{\cal B}^{{\rm I}_1 \cdots {\rm I}_m} ~ \} 
&=& -  \,\, [ ~ {\cal B} \, + \, \fracm 12 \, (m - 2) \, {\cal A}
~] \, R_{\cal A + \cal B}^{{\rm I}_1 \cdots {\rm I}_m} -~ \, [ ~  \fracm 
12 \,{\cal A} \, ({\cal A} + 1  ) ~] \,  U_{\cal A + \cal B}^{{\rm 
I}_1 \cdots {\rm I}_m \, J}{}_J ~~~,\\ \cr 
[ ~ G_{\cal A}{}^{\rm I} \,, \, R_{\cal B}^{{\rm J}_1 \cdots {\rm 
J}_m}~ \} &=&  2 \, (i)^{\s(m)} \,[ ~ {\cal 
B} \, + \, (m - 1) \, {\cal A} \, + \, \fracm 12 \, ~]\, R_{\cal A + 
\cal B}^{{\rm I} \, {\rm J}_1 \cdots {\rm J}_m} \cr 
&&~~~~\,-\, (i)^{\s(m)} \, 
\sum_{r=1}^m \, (-1)^{r -1} \, \d^{I \,  {\rm J}_r} \, R_{\cal A + 
\cal B}^{{\rm J}_1 \cdots {\rm J}_{r-1} \,  {\rm J}_{r+1} \cdots {\rm 
J}_m} \cr  
&&~~~~\,-\,  (-i)^{\s(m)} \, [~{\cal A} \, +\, \fracm 12 ~] \, U_{\cal
A  + \cal B}^{{
\rm J}_1 \cdots {\rm J}_m \, {\rm I}} \cr
&&~~~~\,+\, 2  \, (i)^{\s(m)} \, 
[~ {\cal A}^2 \, -\, \fracm 14 ~] \, U_{\cal A  + \cal B}^{I\, {\rm 
J}_1 \cdots {\rm J}_m \, K}{}_{\rm K} ~~~, \hskip30pt m\ne 2 \\ \cr
[ ~ G_{\cal A}{}^{\rm I} \,, \, R_{\cal B}^{{\rm J}_1 {\rm J}_2}~ \} &=&   
2 ({\cal A} + {\cal B} + \frac 12 ) R_{\cal A+B}^{\rm I J_1 J_2}
-({\cal A}- \frac 12) \, U_{{\cal A+B}}^{\rm J_1 J_2 I} + 2({\cal A}^2
- \frac 14)\, U_{{\cal A + B} \,\,{\rm K}}^{\rm I J_1 J_2 K}\cr
&-& \d^{I J_1} \, \big( \frac 12 G_{\cal A + B}^{\rm J_2} - U_{\cal A
  +B}^{\rm J_2} + 2({\cal A + B} + \frac 12) U_{{\cal A +
    B}\,\,K}^{{\rm J_2 K}} \big) \cr
&+&  \d^{I J_2} \, \big( \frac 12 G_{\cal A + B}^{\rm J_1} - U_{\cal A
  +B}^{\rm J_1} + 2({\cal A + B} + \frac 12) U_{{\cal A +
    B}\,\,K}^{{\rm  J_1 K}} \big) \\ \cr
[ ~ G_{\cal A}{}^{\rm I} \, , \, U_{\cal B}^{{\rm J}_1 \cdots {\rm J}_m}
~ \} &=&  2 \, (i)^{\s(m)} \, [ ~ {\cal B} 
\, + \, (m - 2) \, {\cal A} ~] \, U_{\cal A  + \cal B}^{{\rm I} \, {\rm
J}_1 \cdots {\rm J}_m} \cr
&&~~~~\,- \,2 \, (-i)^{\s(m)} \, [ ~ {\cal A} \, + \, \fracm 12 ~] \, 
\d^{{\rm I} \, {\rm J}_m} \, U_{\cal A  + \cal B}^{ {\rm J}_1 \cdots 
{\rm J}_{m - 1} K} {}_K \cr
&&~~~~\,-\,(i)^{\s(m)} 
\, \sum_{r=1}^{m - 1} \, (-1)^{r -1} \, \d^{{\rm I} \, {\rm J}_r} \, 
U_{\cal A + \cal B}^{{\rm J}_1 \cdots {\rm J}_{r-1} \, {\rm J}_{r+1} 
\cdots  {\rm J}_m} \cr 
&&~~~~\, +\, 2 \,(-i)^{\s(m)}\, \d^{{\rm I} \, {\rm J}_m} \, R_{\cal A
+
\cal  B}^{{\rm J}_1 \cdots {\rm J}_{m-1} } ~~~,\hskip20pt m\ne 2 \\
\cr
[ ~ G_{\cal A}{}^{\rm I} \, , \, U_{\cal B}^{{\rm J}}~ \} &=& -2 i \,
 \d^{\rm I J} ( L_{\cal A + B} + \frac 12({\cal B - A}) \, U_{{\cal A + B} \, {\rm K}}^{\rm
  K} )\, - 2 i \,({\cal A + B} + 1) \, U_{\cal A
  + B}^{\rm I J} \\ \cr
[ ~ G_{\cal A}{}^{\rm I} \, , \, U_{\cal B}^{{\rm J_1} {\rm J_2}}~ \} &=& 
- U_{\cal A+ B}^{\rm J_2} \, \d^{\rm I J_1} + 2 {\cal B} \, U_{\cal A
  + B}^{\rm I J_1 J_2} + \d^{\rm J_2 I} \, \big( G_{\cal A + B}^{\rm J_1} - U_{\cal A +
  B}^{\rm J_1} + 2 {\cal B} \, U_{{\cal A + B} \, {\rm K}}^{\rm J_1 K}\big)
\\ \cr
[ ~ R_{\cal A}^{{\rm I}_1 \cdots {\rm I}_m} \,, \, R_{\cal B}^{{\rm 
J}_1 \cdots {\rm J}_n} ~ \} &=& -\, (i)^{\s(m n)}  \, [ ~ {\cal A} \, -
\, {\cal B} \, - \, \fracm 
12 (m \, - \, n ) ~] \, R_{\cal A + \cal B}^{{\rm I}_1 \cdots {\rm 
I}_m \, {\rm J}_1n \cdots {\rm J}_n} ~~~, \\ \cr
[ ~ R_{\cal A}^{{\rm I}_1 \cdots {\rm I}_m} \,, \, U_{\cal B}^{{\rm J}_1 
\cdots {\rm J}_n} ~ \} &=& (-i)^{\s(m n)} 
\, \sum_{r=1}^m \, (-1)^{r -1} \,\d^{{\rm 
I}_r \, {\rm J}_n} \, R_{\cal A + \cal B}^{{\rm J}_1 \cdots {\rm J}_{n-1
} {\rm I}_1 \cdots {\rm I}_{r-1}  \, {\rm I}_{r+1} \cdots {\rm I}_m}  
\cr
&+& \, i (i)^{\s(m n)} \, [ ~ {\cal B} \, - \, \fracm 12 (n - 2)
~] \, U_{\cal A  
+ \cal B}^{{\rm I}_1 \cdots {\rm I}_m  \, {\rm J}_1 \cdots {\rm J}_n }, \hskip10pt m \ne 2, \, n \ne 1 \\ \cr
[ ~ R_{\cal A}^{{\rm I}_1 {\rm I}_2} \,, \, U_{\cal B}^{{\rm
    J}} ~ \} &=& ({\cal B} + \frac 12) U_{\cal A + B}^{\rm I_1 I_2 J}
\cr
&+& \frac 12 \big( U_{\cal A + B}^{\rm I_2} - G_{\cal A + B}^{I_2} + 2
({\cal A + B} + \frac 12)\, U_{{\cal A+B}\,\,{\rm K}}^{\rm I_2 K}
\big) \d^{\rm J_1 I_1} \cr
&-& \frac 12 \big( U_{\cal A + B}^{\rm I_1} - G_{\cal A + B}^{I_2} + 2
({\cal A + B} + \frac 12)\, U_{{\cal A+B}\,\,{\rm K}}^{\rm I_1 K}
\big) \d^{\rm J_1 I_2} \\ \cr
[ ~ U_{\cal A}^{{\rm I}_1 \cdots {\rm I}_m} \,, \, U_{\cal B}^{{\rm J}_1 
\cdots {\rm J}_n} ~ \} &=&-  \,  (i)^{\s(m n)}\, \Big\{ \, \sum_{r =
1}^m \, (-1)^{r-1} 
\, \d^{{\rm I}_m {\rm J}_r} \, \, U_{\cal A + \cal B}^{{\rm I}_1 \cdots 
{\rm I}_{m-1} \, {\rm J}_1 \cdots {\rm J}_{r-1} \, {\rm J}_{r+1} \cdots 
{\rm J}_{n-1} {\rm J}_n }~ \cr
- (-1)^{mn}&\sum_{r = 1}^m &\, 
(-1)^{r-1} \, \d^{{\rm I}_r {\rm J}_n} \, \, U_{\cal A + \cal B}^{{\rm 
J}_1 \cdots {\rm J}_{n-1} \, {\rm I}_1 \cdots {\rm I}_{r-1} \, {\rm 
I}_{r+1} \cdots {\rm I}_{m-1} {\rm I}_m } \, \Big\}, \hskip10pt n \ne
1 \\ \cr
[ ~ U_{\cal A}^{{\rm I}_1} \,, \, U_{\cal B}^{{\rm J}_1 
\cdots {\rm J}_m} ~ \} &=& - (i)^{\sigma(m)} \, \{ \sum_{r=1}^n
(-1)^{r-1}\, \d^{\rm I_1 J_r} \, U_{\cal A + B}^{\rm J_1 \cdots {\hat
    J_r} \cdots J_n}.
\end{eqnarray}
where the function $\s(m)=0$ if $m$ is even and $-1$ if $m$ is odd.
The central extensions $c$ and ${\tilde c}$ are unrelated since
we have only imposed the Jacobi identity.

\subsection{Short Distance Expansion for $D(y)\,O(x)$}

~~~~The transformation laws for $\xi(y)$ on the dual of the algebra
allows us to extract the short distance expansion rules  for the field 
$D(y)$ and any other element in the coadjoint representation $O(x)$.
Using the transformation rules
\begin{eqnarray}
L_\xi *( {\bar L}_{D},{\bar \b}) &=&~ {\bar L}_{{\tilde D}}
 ~~~,~~~
{\tilde D} ~=~ -2 \,\xi' D \,-\, \xi \, D' \,-\, \fracm {c {\bar 
\beta}} 8 \, \xi''' \,\d_{N,2} ~~~, \\ \cr
L_\xi * {\bar G}^{\bar Q}_{\Psi^{\bar Q}} &=&~  {\bar G}^{\bar Q}_{
\tilde {\Psi^{\bar Q}}} ~~~~,~~~ {\tilde {\Psi^{\bar Q}}} ~=~ 
-(\fracm 32 \xi' \psi^{\bar Q} + \xi (\psi^{\bar Q})') ~~~,\\ \cr
L_\xi * {\bar T}^{\rm{\bar R} {\bar S}}_{\tau^{\rm{\bar R}{\bar S}}}  
&=&~ {\bar T}^{\rm{\bar R} {\bar S}}_{{\tilde \tau}^{\rm{\bar R}{\bar 
S}}} ~~~,~~~ {\tilde \tau}^{\rm{\bar R}{\bar S}} \,=\, - \xi'
{\tau^{\rm{\bar R}{\bar S}}} \,-\, \xi \, ({\tau^{\rm{\bar R}{\bar 
S}}})' ~~~,
\\ \cr
L_\xi * {\bar U}^{\rm {\bar V_1} \cdots {\bar V_n}}_{\o^{\rm {\bar
V_1} \cdots {\bar V_n}}} &=&~ {\bar U}^{\rm {\bar V_1} \cdots {\bar 
V_n}}_{{\tilde \o}^{\rm {\bar V_1} \cdots {\bar V_n}}} \,+\, \fracm i2
(i)^{[\fracm{n-2}{2}]-[\fracm n2 ]}\, {\bar R}^{\rm [{\bar V_1} \cdots
{\bar V_{n-2}}}_{\xi''{\o^{\rm {\bar  V_1} \cdots {\bar V_n}}}}
\d^{\rm V_{n-1}] V_{n}} ~~~,\label{LU} \\ \nonumber \cr
{\rm where \,\,\,\,\,\,\,\,}{{\tilde \o}^{\rm {\bar V_1} \cdots {\bar 
V_n}}} \,&=&\, (\fracm n2 -2) \, \xi' \, {\o^{\rm {\bar V_1} \cdots 
{\bar V_n}}} \,-\,  \xi ({\o^{\rm {\bar V_1} \cdots {\bar V_n}}})' 
~~~,  \\\cr
L_\xi * {\bar R}^{\rm {\bar T}_1 \cdots {\bar T}_m}_{\rho^{\rm {\bar
T_1} \cdots {\bar T_m}}} &=&~ {\bar R}^{\rm {\bar T_1} \cdots {\bar
T_m}}_{{\tilde \rho}^{\rm {\bar T_1} \cdots {\bar T_m}}} ~~~,~~~ \\ \cr 
{\rm where \,\,\,\,\,\,\,} {{\tilde \rho}^{\rm {\bar T_1} \cdots {\bar 
T_m}}}&=& (\fracm {\rm m}2 -2) \, \xi' \, \rho^{\rm {\bar T_1} \cdots 
{\bar T_m}} - \xi  \, (\rho^{\rm {\bar T_1} \cdots {\bar T_m}})' \nonumber ~~~
\end{eqnarray}
and $[x]$ is the greatest integer in $x$.

By constructing the generators we  have the short distance expansion laws,
\begin{eqnarray}
D(y) D(x)  &=& {-1 \over 2 \pi (y-x)}\, \partial_x D(x) - {1 \over
  \pi i (y-x)^2}\, D(x) - {3 \,c\,\b \,\d_{N,2}\over 4 \pi i(y-x)^4} \\ \cr
D(y) \, \psi^Q(x) &=& {-3 \over 4 \pi i  (y-x)^2} \, \psi^Q(x) - {1 
\over 2  \pi i (y-x)} \partial_x \psi^Q(x)\\ \cr
D(y) \, A^{R S}(x) &=& {-1 \over 2 \pi i  (y-x)^2} \, A^{R S}(x) - 
{1 \over 2 \pi i (y-x)} \partial_x A^{R S}(x)\\ \cr
D(y) \, \o^{V_1 \cdots V_n}(x) &=& {n-4\over 4 \pi i  (y-x)^2} 
\,\o^{V_1 \cdots V_n}(x) - {1 \over 2 \pi i (y-x)}\partial_x \o^{V_1 
\cdots V_n}(x)\\ \cr
D(y) \, \r^{V_1 \cdots V_n}(x) &=& {n-4\over 4 \pi i  (y-x)^2} \,\r^{V_1 
\cdots V_n}(x) - {1 \over 2 \pi i (y-x)}\partial_x \r^{V_1 \cdots V_n}
(x) \cr\cr
&+& {(i)^{[\frac n2 ]- [ \frac{n-2}2 ]} \over 2 \pi (y-x)^3 } \o^{I_1 
\cdots I_n A}_A(x) \label{DR}
\end{eqnarray}

It is worth commenting on the last summand that appears in
Eq.(\ref{DR}).  From Eq.(\ref{LU}), we notice that  the transformation
of $\o^{V_1 \cdots V_n}(x)$ contributes to $\r^{I_1 \cdots
  I_m}(x)$. This implies that the charge $Q_\xi$ contains the summand
\begin{equation}
Q_\xi = \cdots + \,\,i (i)^{[\frac{q-2}2] - [\frac q2 ]} \frac 12
\int\, \Pi_{\r}^{I_1 \cdots  I_{q-2}}(x) \left(  \xi''(x) \o^{I_1 \cdots
    I_{q-2} I_{q-1} I_q }(x)\,  \right) \d_{{q-1}, q}\, dx ,
\end{equation}
where $\Pi_{\r}^{I_1 \cdots  I_{q-2}}(x)$ is the conjugate momentum 
for ${\r}^{I_1 \cdots  I_{q-2}}(x)$. So 
\begin{equation}
\{ Q_\xi , \r^{I_1 \cdots I_{q-2}}(x) ] = \xi(x) {\d \over \d
  \xi(x) } \r^{I_1 \cdots I_{q-2}}(x),
\end{equation}
will lead to 
\begin{equation}
D(y) \, \r^{I_1 \cdots I_n}(x) = \frac i2  i^{[\frac n2 ]-[ \frac{n-2}2
  ]}\, \partial_y^2 \d(y,x)\, \o^{I_1 \cdots I_n A}_A(x) +  ({\rm
  plus}\,  \r^{I_1
  \cdots I_n} \, {\rm dependent \,\, terms}).
\end{equation}
The existence of contributions like $\xi ''$  as seen in Eq.(\ref{LU}) is
due to the transformation of the connection $\G$.  Although the connection 
does not appear in the algebra since $g_{ a b} = 1$, its presence is 
felt in the coordinate transformations.

\subsection{Short Distance Expansion for $\psi^I(y)\,O(x)$}
~~~~The next sections follow in the same way as the $D(y) \, O(x)$
expansions.  One constructs the generators dedicated to the symmetry
transformation of the particular algebraic element and then identifies
the short distance expansion or operator product expansion from the
variation.   From \cite{Curto:2000kd} we can write the transformation
laws due to the spin $\frac 12 $ fields as
\begin{eqnarray}
G^{\rm I}_{\chi^{\rm I}} * {\bar G}^{\rm \bar Q}_{\psi^{\rm \bar Q}} 
&=&~ \d^{\rm I {\bar Q}}\, {\bar L}_{\tilde \xi}+ {\bar 
T}^{\rm I {\bar Q}}_{{\tilde A}^{\rm I {\bar Q}} } + (1-\d^N_2)\,\big(
{\bar R}^{\rm I {\bar  Q}}_{\chi^{\rm I} \psi^{\rm {\bar Q}}} - {\bar U}^{\rm I {\bar  Q}}_{\chi^{\rm I} \psi^{\rm {\bar Q}}}\big)\cr \cr
 {\rm where}&&  {\tilde \xi}=\fracm 12 \, (\psi^{\bar Q})' 
\chi^I -\fracm 32 \, (\chi^I)' \psi^{\bar Q}, ~~~{\tilde A}^{\rm I
  {\bar Q}} = (2\chi^{\rm I} \psi^{\rm \bar Q}-2\chi^{\rm Q} \psi^{\rm \bar I} ) \\\cr
G^{\rm I}_{\chi^{\rm I}} * ({\bar L}_D, {\bar \b}) &=&~ 4 i \, {\bar
G}^{\rm I}_{{\tilde \chi}^{\rm I}} - 2 i
{\tilde U}^{\rm I}_{\chi^{\rm I} D} ,  \cr \cr 
{\rm where}&& {\tilde \chi}^{\rm I} = (-\chi^{\rm I}\,
D-{\bar \b} c \,\d^N_2\, (\chi^{\rm I})'')\\ \cr
G^{\rm I}_{\chi^{\rm I}} * {\bar T}^{\rm {\bar R\,S}}_{\tau^{\rm{\bar 
R\,S}}} &=&~ \fracm i2 ({\bar G}{}^{\rm S}_{\chi^{\rm S}} \, \d^{\rm { 
\bar R} I} - {\bar G}{}^{\rm \bar R}_{\chi^{\rm \bar R}}\, \d^{\rm I 
{\bar S}}) ~~~, ~~~ \chi^{\rm{\bar R}} \,=\,\chi^{\rm S} = 2 (\chi^{\rm 
I})' \tau^{\rm{\bar R\,S}} + \chi^{\rm I} \, (\tau^{\rm {\bar R\,
  S}})'  \\ \cr
G^{\rm I}_{\chi^{\rm I}} * {\bar U}^{\rm V_1 V_2}_{\o^{\rm V_1 V_2}}&=& \, -i {\bar U}^{\rm I}_{(-2 {\chi^{\rm I}}' \o^{\rm V_1 V_2}
- {\o^{\rm V_1 V_2}}' \chi^{\rm I})}  \, \d^{\rm V_1 V_2} + 2 i {\bar
U}^{\rm V_2}_{({\o^{\rm V_1 V_2}}' \chi^{\rm I})} \d^{\rm V_1 I},
~~~~~~~~{N \ne 2} \\ \cr
G^{\rm I}_{\chi^{\rm I}} * {\bar R}^{\rm {L M}} &=& -3 i \, {\bar
  R}^{\rm I L M}_{(\chi^{\rm I} \, \r^{\rm LM})} \\ \cr
G^{\rm I}_{\chi^{\rm I}} * {\bar R}^{\rm {\bar T_1} \cdots {\bar
T_m}}_{\rho^{\rm {\bar T_1} \cdots {\bar T_m}}} &=&~ 2 i (i)^{m+1}
(i)^{[\fracm{m+2}{2} ]-[\fracm m2 ]}\, {\bar U}^{\rm [{\bar T_1} 
\cdots {\bar T_m}]}_{(\chi^{\rm I} \r^{\rm {\bar T_1} \cdots {\bar
T_m}})}~~~~~~~~~~~~~m \ne 2\cr
&&\,\,\,\,\,\,\,\,\, - 2 i^{[{m-1\over 2}]-[\fracm {m-2}2]} \, 
\d^{I\,[{\bar T_1}}\,{\bar R}^{\rm {\bar T_2} \cdots {\bar T_m}]}_{
((\chi^{\rm I})' \r^{\rm {\bar T_1} \cdots {\bar T_m}}-(\chi^{\rm 
I}) (\r^{\rm {\bar T_1} \cdots {\bar T_m}})')}\cr
&&\,\,\,\,\,\,\,\,\, - (i) (i)^{[\fracm {m+1}2 ]-[\fracm m2
]}\sum^{m+1}_{r=1}\,(-1)^{r-1}\, {\bar R}^{\rm  {\bar T_1} \cdots 
{\bar T_{r-1}}\, I \,{\bar T_{r+1}}\cdots {\bar T_m}}_{(\chi^{\rm 
I}\r^{\rm {\bar T_1} \cdots {\bar T_m}})}\\ \cr
G^{\rm I}_{\chi^{\rm I}} * {\bar U}^{\rm {\bar V_1}\cdots {\bar
V_n}}_{\o^{\rm {\bar V_1} \cdots {\bar V_n}}} &=&~ - 2 i^{[{n-1\over
2}]-[\fracm n2]} \, \d^{I\,[{\bar V_1}}\,{\bar U}^{\rm {\bar V_2}
\cdots {\bar V_n}]}_{((n-4)(\chi^{\rm I})' \o^{\rm {\bar V_1} \cdots
{\bar V_n}}-(\chi^{\rm I}) (\o^{\rm {\bar V_1} \cdots {\bar
V_n}})')} \cr \cr
&&\,\,\,\,\,\,\,\,\,\, + 2(-1)^{n-1} (i)^{[\fracm {n-1}2]-[\fracm 
n2]} \,\d^{{\rm I}[ {\rm \bar V_n}} {\bar U}^{{\bar V_1}\cdots {\bar 
V_{n-1}}] \,{\rm K}}_{((\chi^{\rm I})' \o^{\rm {\bar V_1} \cdots {\bar
V_n}})}{}_{\rm K} \cr \cr
&&\,\,\,\,\,\,\,\,\,\, + (i) (i)^{[\fracm {n-1}2 ]-[\fracm n2
]}\sum^n_{r=1}\, {\bar U}^{\rm [ {\bar V_1} \cdots {\bar V_{r-1}}{
\bar V_{r+1}} \cdots {\bar V_n}}_{(\chi^{\rm I}\o^{\rm {\bar V_1} 
\cdots {\bar V_n}})}\,\d^{\rm{\bar V_r}\,]\,I}\cr \cr
&& \,\,\,\,\,\,\,\,\,\,+ {\bar G}^{ [ {\rm \bar V_2}}_{( - 4 i 
\, (\chi^{\rm I})' (\o^{\rm {\bar V_1}\cdots {\bar V_n}})'-2i\,
(\chi^{\rm I}) (\o^{\rm {\bar V_1}\cdots {\bar V_n}})'')} \,\d^{\rm 
\bar {V_3} \bar{V_4}}\, \d^{{\rm \bar V_1} ]{\rm I}}\, \d^{n 4} \cr \cr
&&\,\,\,\,\,\,\,\,\,\, - (-1)^{n-1} \, \d^{\rm I [ {\bar V_n}}\, {\bar R}_{((\chi^{\rm I})' \o^{\rm {\bar V_1} \cdots {\bar V_n}})}^{\rm {\bar V_1}\cdots {\bar V_{n-1}}]} \cr\cr
&&\,\,\,\,\,\,\,\,\,\, -2\, \d^{\rm I [ {\bar V_1}}\, {\bar R}_{((\chi^{\rm I})'' \o^{\rm {\bar 
V_1} \cdots {\bar V_n}})}^{\rm {\bar V_2}\cdots {\bar V_{n-2}}}
\,\d^{\rm V_{n-1}] V_n} \,\,\,\, n \ne 1,2 \\ \cr
G^{\rm I}_{\chi^{\rm I}} * {\bar U}^{\rm {\bar V_1}}_{\o^{\rm {\bar
      V_1}}}&=& - {\bar R}^{\rm I V_1}_{(\chi^{\rm I}\, \o^{\rm K})} -
{\bar U}^{\rm I K}_{(\chi^{\rm I} \o^{\rm K})} \,+\, {\bar U}^{\rm K
  I}_{(\o^{\rm K}\, \chi^{\rm I} )}
\end{eqnarray}

The short distance expansions that follow from here are:
\begin{eqnarray}
\psi^I(y)\, D(x) &=& {- i \over 4 \pi (y-x)} \, \partial_x (\psi^I
(x)) - {3\over 4 \pi i (y-x)^2} \, \psi^I(x) \\ \cr
\psi^I(y)\,\psi^Q(x) &=& {-4 i \over (y-x)}\,
\d^{I Q}\,D(x) -\,\d^N_2 \,{8 i \b c \over (y-x)^2} \d^{I Q} \\ \cr
\psi^A(y)\,A^{R S}(x) &=& {\pi \over i (y-x)}\,\left(\d^{A R}\d^{L 
S} - \d^{A S}\d^{L R} \right)\, \psi^L(x)\\ \cr
\psi^J(y) \,\o^{A_1 \cdots A_{q-1}}_{A_q}(x) &=&
(1-\d^{q}_2 ){1 \over 2 \pi i (y-x)}2(i)(i)^q\,(i)^{[\frac{q+1}2]-[\frac{q-1}2]}\,
\d^J_{A_q}\, \r^{A_1 \cdots A_{q-1}}(x)\cr \cr
&-& { i^{[\frac q2 ]-[ \frac{q+1}2]} \over  \pi i (y-x)} \left( 
{(q-3)\over (y-x)} \o^{J A_1 \cdots A_{q-1}}_{A_q}(x) - \partial_x \o^{J A_1
  \cdots A_{q-1}}_{A_q}(x)\right) \cr \cr
&+&(1-\d^{q}_2)(-1)^{(q-2)} \, (-i)^{[\frac {q-2}2 ]- [\frac{q-1}2]}{1 \over \pi i
    (y-x)^2} \,\o^{[A_1 \cdots A_{q-2} J}\d^{A_{q-1}]}_{A_q}(x) \cr \cr
+ \,\,\,{(i)^{[\frac{q-1}2] - [\frac q2]} \over 2 \pi i (y-x)}
&{\displaystyle \sum_{r=1}^q} & \,\d^{A_1}_{[V_{1}} \cdots
\d^{A_{r-1}}_{V_{r-1}}\d^{I}_{V_{r}}\cdots \d^{A_{q-1}}_{V_q}
\d_{V_{q+1}] A_q }\, \o^{V_1 \cdots V_{q+1}}(x)\cr\cr
&-& {  i \d^q_3 (1- \d^N_2) \over 2 \pi (y-x)}  \, \d^{\rm J [A_1} \,
\psi^{\rm A_2]}(x) \cr
&-& i {\d^q_2 (1- \d^N_2) \over 2 \pi (y-x)}\d^{\rm J [ A_1}\, \o^{\rm A_2]}(x) \\ \cr
\psi^J(y)\, \r^{A_1 \cdots A_q}(x) &=& {- (-1)^{(q+1)} \over 2 \pi
  (y-x)} (i)^{[\frac{q+1}{2}] - [\frac{q}{2}]} \o^{A_1 \cdots A_q J}(x)
+{ i \d^q_2 (1- \d^N_2) \over 2 \pi (y-x)}  \, \d^{\rm J [A_1} \, \psi^{\rm A_2]}(x)
\cr \cr
{- (-1)^{(q-1)} \over 2 \pi  (y-x)} (i)^{[\frac{q+1}{2}] -
[\frac{q}{2}]}
&{\displaystyle \sum_{r=1}^{q+1}} & \,\d^{A_1}_{[V_{1}} \cdots
\d^{A_{r-1}}_{V_{r-1}}\d_{I}^{A_{r}}\d_{V_{r}}^{A_{r+1}}\cdots 
\d^{A_{q}}_{V_{q-1}]} \,\d^{I J}\, \r^{V_1 \cdots V_{q-1}}(x)(1-\d^q_2) \cr\cr 
&-& { i^{[\frac{q-1}2 ] - [ \frac{q-2}2 ]} \over \pi i (y-x)} 
\left( {1 \over (y-x)} \, \r^{J A_1 \cdots A_q} - \partial_x 
\r^{J A_1 \cdots A_q} \right) \cr \cr
&-& i {\d^q_2 (1- \d^N_2) \over 2 \pi (y-x)}\d^{\rm J [ A_1}\, \o^{\rm A_2]}(x) 
\end{eqnarray}
\subsection{Short Distance Expansion for $A^{J K}(y)\, O(x)$}
As a generator of SO(N) transformations, the operator $A^{ J K}(y)$
takes on its own significance from the symmertic part of $U^{ J K}(y)$.  Here are some of the expansions for 
operators $O(x)$ paired with $A^{J K}(y)$.
\begin{eqnarray}
T^{\rm J\, K}_{t^{\rm J\, K}} * {\bar G}^{\rm \bar Q}_{\psi^{\rm \bar 
Q}} &=&~ -2 ({\bar G}^{\rm K}_{(t^{\rm J\, K}\, \psi^{\rm \bar Q})} 
\d^{\rm {\bar Q} J}-{\bar G}^{\rm J}_{(t^{\rm J\, K}\, \psi^{\rm \bar 
Q})} \d^{\rm {\bar Q} K}) ~~~,\\ \cr 
T^{\rm J\, K}_{t^{\rm J\, K}} * {\bar U}^{\rm {\bar V_1}\cdots {\bar
V_n}}_{\o^{\rm {\bar V_1} \cdots {\bar V_n}}} &=&~ -\sum_{r=1}^{n-1} 
(-1)^{r+1} ( \d^{\rm J\, [ \, {\bar V_1}}\, {\bar U}^{\rm {\bar V_2}
\cdots {\bar V_{r-1}} | \,K\,| {\bar V_{r+1}} \cdots]{\bar V_n}}_{
(t^{\rm J\, K}\,\o^{\rm {\bar V_1} \cdots {\bar V_n}})}\cr \cr 
&-& \d^{\rm K\, [ \, {\bar V_1}}\, {\bar U}^{\rm {\bar V_2}\cdots {\bar V_{r-1}} 
| \,J\,| {\bar V_{r+1}} \cdots]{\bar V_n}}_{(t^{\rm J\, K}\,\o^{\rm 
{\bar V_1} \cdots {\bar V_n}})}) \cr \cr
&+& {\bar U}^{\rm [{\bar V_1}\cdots {\bar V_{n-1}}]\,J\,}_{
(t^{\rm J\, K}\,\o^{\rm {\bar V_1} \cdots {\bar V_n}})}\,\d^{\rm {\bar 
V_n}\,K} -{\bar U}^{\rm [{\bar V_1}\cdots {\bar V_{n-1}}]\,K\,}_{(t^{
\rm J\, K}\,\o^{\rm {\bar V_1} \cdots {\bar V_n}})}\,\d^{\rm {\bar
V_n}\,J} \cr \cr
&-i& (-1)^{n-2}\,(\d^{\rm K\, {\bar V_n}}\, \d^{J \,[ {\bar 
V_1}}\,-\, \d^{\rm J\, {\bar   V_n}}\,\d^{\rm K\,[{\bar V_1}}) {\bar 
R}^{\rm {\bar V_2}\cdots {\bar V_{n-1}}]}_{((t^{\rm J\, K})'\,\o^{\rm 
{\bar V_1} \cdots {\bar V_n}})} ~~~,\\ \cr
T^{\rm J\, K}_{t^{\rm J\, K}} * {\bar R}^{\rm {\bar T_1}\cdots {\bar
T_m}}_{\r^{\rm {\bar T_1}\cdots {\bar T_m}}} &=&~ \sum_{r=1}^{m}
\,(-1)^{r+1}\,(\d^{\rm [ {\bar T_1} |\,J\,|} \,{\bar R}^{\rm {\bar
T_2}\cdots {\bar T}_{r-1} \,|\,K\,|\,{\bar T_{r+1}}\cdots {\bar
T_m}]}_{(t^{\rm J\, K}\,\r^{\rm {\bar T_1} \cdots {\bar T_m}})}\cr \cr
&&  -\d^{\rm [ {\bar T_1} |\,K\,|} \,{\bar R}^{\rm {\bar T_2} \cdots {\bar 
T}_{r-1} \,|\,J\,|\,{\bar T_{r+1}} \cdots {\bar T_m}]}_{(t^{\rm J\, 
K}\,\r^{\rm {\bar T_1} \cdots {\bar T_m}})}) ~~~,
\\ \cr 
T^{\rm J\, K}_{t^{\rm J\, K}} * ({\bar T}^{\rm {\bar R}{\bar S}}_{
\tau^{\rm {\bar R}{\bar S}}}, {\bar \b}) &=&~ \fracm 12 (\d^{\rm {\bar 
R}J} \d^{\rm {\bar S} K} -  \d^{\rm {\bar R} K} \d^{\rm {\bar S} J})\, 
{\bar L}_{((t^{\rm J K})' \, \tau^{\rm {\bar R}\,{\bar S}})} 
+  \fracm 12 {\bar T}^{\rm A\,B}_{(t^{\rm J\, K}\, \tau^{\rm {\bar 
R}{\bar S}})} \, \d^{\rm J K {\bar R} {\bar S}}_{\rm A B} + 4 {\bar 
\b} {\bar T}^{\rm J K}_{(\t^{\rm {\bar R}{\bar S}})'} ~~~,\cr \cr 
{\rm where} \,\,\,\,\,\,\d^{\rm J K {\bar R} {\bar S}}_{\rm A B} 
&\equiv& (\,{ \d^{\rm A K} \d^{\rm B {\bar S}} \d^{\rm {\bar R}
J} - \d^{\rm A K} \d^{\rm B {\bar R}} \d^{\rm {\bar S} J} + \d^{
\rm A {\bar S}} \d^{\rm B J} \d^{\rm {\bar R} K} } \cr  \cr
&& - \d^{\rm A {\bar R}} \d^{\rm J S} 
\d^{\rm {\bar S} K} + {\rm \d^{A {\bar S}} \d^{\rm B K} \d^{\rm 
{\bar R}J} - \d^{\rm A {\bar R}} \d^{\rm K B} \d^{\rm {\bar S} J} 
}\cr \cr
&& + \d^{\rm A J} \d^{\rm B {\bar S}} \d^{\rm { \bar R} K} - \d^{\rm J
A} \d^{\rm {\bar R} B} \d^{\rm {\bar S} K} + {\rm \d^{A {\bar S}} 
\d^{B K} \d^{{\bar R} J}  } \, )~~~,
\end{eqnarray}

We get the $A^{J K}(y)\, D(x)$ short distance expansion as:
\begin{eqnarray}
A^{J K}(y)\,D(x)& =& {1 \over 4 \pi i (y-x)^2}\,  \left( \d^{R S} 
\d^{J K} - \d^{R K} \d^{S J} \right) \, A_{R S}(x) \\ \cr
A^{A B}(y) \, \psi^C(x)& = &{-1 \over \pi i (y-x)} \left( \d^{A C}
  \psi^B(x) - \d^{B C} \psi^A(x)\right) \\ \cr
A^{J K}(y)\,A^{R S}(x)& = &{1\over 4 \pi i (y-x)} \d^{JKRS}_{AB} 
\, A^{A B}(x) + {\b\over \pi i (y-x)^2}\left( \d^{J R} \d^{K S} 
- \d^{R K} \d^{S J} \right) \\ \cr
A^{J K}(y) \, \o^{A_1 \cdots A_n}(x)&=& {\displaystyle \sum_{r=1
}^{n-1}} {(-1)^{r+1}\over 2 \pi (y-x)} \left( \d^{K A_r} \, {\o}^{J 
A_1 \cdots {\hat A_r} A_{n-1} A_n}(x) - \d^{J A_r} {\o}^{K A_1 
\cdots {\hat A_r} A_{n-1} A_n}(x) \right) \cr \cr
&+& {1\over 2 \pi i (y-x)} \left( \d^{A_n J} \, \o^{A_1 \cdots 
A_{n-1} K}(x) - \d^{A_n K} \,  \o^{A_1 \cdots A_{n-1} J}(x) \right)  
\\ \cr
A^{J K}(y) \, \r^{A_1 \cdots A_n}(x)&=&{\displaystyle \sum_{r=1}^{n}} 
{(-1)^{r+1}\over 2 \pi (y-x)} \left( \d^{K A_r} \, {\r}^{J A_1 \cdots 
{\hat A_r} A_{n-1} A_n}(x) - \d^{J A_r}{\r}^{K A_1 \cdots {\hat A_r} 
A_{n-1} A_n}(x) \right) \cr \cr
&+& { i (-i)^{n-1}\over 2 \pi i (y-x)^2} \left( \o^{ J A_1 \cdots A_n  
K}(x) - \o^{K A_1 \cdots A_n  J}(x) \right)
\end{eqnarray}
\subsection{Short Distance Expansion for $\o^{I_1 \cdots I_q}(y) \,
  O(x)$}
~~~~The next two sections provide the short distance expansions for the
fields $\o^{I_1 \cdots I_q}$ and $\r^{P_1 \cdots P_q}$.  These fields
appear in the  ${\cal GR}$ realization as natural Grassmann extensions
of the SO(N) gauge field and the one dimensional diffeomorphism
field. The relevant transformation laws are :
\begin{eqnarray}
{U}^{\rm J}_{\m^{\rm J}} * {\bar L}_{D} &=& -2\,i \, {\bar G}^{\rm
  J}_{(\m^{\rm J} \, D)} \\ \cr
U^{\rm J}_{\m^{\rm J}} * {\bar U}^{\rm K}_{\o^{\rm K}},&=& {\bar
  R}^{\rm J K}_{(\m^{\rm J} \o^{\rm K})} \\ \cr
U^{\rm J}_{\m^{\rm J}} * {\bar G}^{\rm K}_{\psi^{\rm K}} &=& - {\bar
  R}^{\rm J K}_{(\m^{\rm J} \, \psi^{\rm K})} \\ \cr
U^{\rm J}_{\m^{\rm J}} * {\bar U}^{\rm L M N}_{\o^{\rm L M N}} &=&
{\bar R}^{\rm L M}_{(\m' \o)} \, \d^{\rm J N} + 4 {\bar R}^{\rm J
  L}_{(\m \o')} \, \d^{\rm M N} \\ \cr
U^{\rm J_1 J_2}_{\m^{J_1 J_2}} * {\bar U}^{\rm K}_{\o^{\rm K}} &=&
{\bar G}^{\rm J_1}_{(\o^{\rm J_1 J_2} \, \o^{\rm K})} \, \d^{\rm K \,
  J_2} \, - \, {\bar G}^{\rm J_2}_{(\m^{\rm J_1 J_2} \, \o^{\rm K})}
\, \d^{\rm K J_1} \\ \cr
U^{\rm J_1 J_2}_{\m^{\rm J_1 J_2}} * {\bar G}^{\rm K}_{\psi^{\rm K}}
&=& - \d^{\rm J_1 K} \, {\bar G}^{\rm J_2}_{(\u^{\rm J_1 J_2}\,
  \psi^{\rm K})} \\ \cr
U^{\rm I_1 \cdots I_q}_{\mu^{\rm \{I_q\}}}*{\bar U}^{\rm {\bar V_1} 
\cdots {\bar V_m}}_{\o^{\rm \{V_m\}}} &=& -\d^{\rm m\,q}\,\d^{\rm 
[\,I_1 \cdots I_q\,]}_{\rm [{\bar V_1}\cdots {\bar V_q}\,]} {\bar 
L}_{((\fracm {4-q}2) \mu' \o-(\fracm {q-2}2) \mu \o')}\, - 2\, {\bar 
T}^{\rm {\bar V_m} I_q}_{(\mu \o)} \d^{\rm [\, I_1 \cdots I_{q-1}\,]}_{\rm 
[{\bar V_1}\cdots {\bar V_{q-1}}\,]}\,\d^{\rm q}_{\rm m} \cr \cr
&-& 2i^{([\fracm {q}2]-[\fracm {q+1}2])}\,
\d^{\rm m, (q+1)} {\bar G}^{\rm [ {\bar V_1}}_{(-(q-2)\mu \o'+(3-q)\mu'\o ) 
}\, \d^{\rm {\bar V_2} \cdots {\bar V_m}\,]}_{\rm [ I_1 \cdots I_q ]}
\cr \cr
&+& 2 (-1)^q (i)^{[\fracm q2 ]-[\fracm {q+1}2]} {\bar G}^{\rm
  I_q}_{(- \mu\o)'}\,\d^{\rm m,q+1} \, \delta^{\rm [ {\bar V_1}\cdots 
{\bar V_{m-2}}}_{\rm [I_1 \cdots I_{q-1}]}\, \d^{\rm {\bar V_{m-1}}], 
{\bar V_m}}\cr \cr
&-& i (i)^{[\fracm q2 ]-[ \fracm {q-1}2]}\sum_{r=1}^{q-1} (-1)^{
r-1} \d^{\rm m}_{\rm q-1}\, {\bar G}^{\rm [ I_r}_{(\mu \o)} \d_{\rm \bar 
[ V_1}^{\rm I_1}\cdots \d_{\rm \bar V_{r-1}}^{\rm I_{r-1}} \, \d_{\rm 
\bar V_r}^{\rm I_{r+1}}\cdots \d_{\rm \bar V_m]}^{\rm I_q ]}
\cr\cr
&+&\sum_{r=1}^{q-1} 2 (-1)^{r+1} ( {\bar T}^{\rm I_r [{\bar V_1}
}_{(\mu \o)} \d_{\rm [ I_{1}}^{{\bar V}_2} \cdots  \d_{{\rm I_{r-1}
}}^{{\bar V}_r} \d_{{\rm I_{r+1}}}^{{\bar V}_{r+1}} \cdots \d_{{\rm 
] I_{q}}}^{{] \bar V}_m} \d^{\rm q,m})\cr \cr
&+& i (i)^{\{[\fracm q2]+[\fracm {m-q}2 +2]-[\fracm {m+2}2]\}}
\cr\cr
&\times& \{ \sum_{r=1}^{m-q}(-1)^{r-1}\d_{\rm  [\,I_1 \cdots I_{q-1}\,]
}^{\rm  [{\bar V_1}\cdots {\bar V_{q-1}}\,}\ {\bar U}^{\rm {\bar V_q} 
\cdots {\bar V_{q+r-1}} I_q {\bar V_{q+r}}\cdots {\bar ] V_m}}_{\mu \o} 
\cr\cr
&-&(-1)^{q(m-q+2)}\sum_{r=1}^q (-1)^{r-1} {\bar U}^{\rm {\bar
V_1}
\cdots {\bar V_{m-q+1}} [\, I_r}_{\mu\o }  \d^{\rm I_1 \cdots I_{r-1}
I_{r+1}\cdots \, I_q ]}_{\rm {\bar V_{m-q+2}}\cdots{\bar V_{m-q+2+r}}
\cdots{\bar V_m}}~\}\cr \cr
&-&(i)^{\{[\fracm q2] +[\fracm {m-q}2] -[\fracm {q+m-4}2]\}}
\, {\bar R}^{\rm [{\bar V_1} \cdots {\bar V_{m-q}}}_{(\mu \o)'} \, 
\d_{\rm [\,I_1 \cdots ] I_{q}\,}^{\rm {\bar V_{m-q+1}\cdots ] {\bar
V_{m}}\,}} ~~~, \\\cr
U^{\rm I_1 \cdots I_q}_{\mu^{\rm \{I_q\}}} *{\bar R}^{\rm {\bar 
T_1} \cdots {\bar T_m}}_{\r^{\rm \{T_m\}}} &=& -i (-1)^{q(m-q+2)}
(i)^{\{[\fracm{m-q}2+2]+[\fracm q2]-[\fracm m2]\}}\,\cr\cr
&\times& \sum_{r=1}^{m-q+2}\, (-1)^{r-1}\d^{\rm [\,I_1 
\cdots I_{q-1}\,] }_{\rm [{\bar T_{1}\cdots {\bar T_{q-1}}\,]}} 
{\bar R}^{\rm {\bar T_q} \cdots {\bar T_{q+r-1}\, I_q \,{\bar 
T_{q+r+1}\cdots {\bar T_{m}}}}}_{\mu \r}
\end{eqnarray}
For $q > 2$ the short distance expansions are:
\begin{eqnarray}
\o^{I_1 \cdots I_q}(y) \, D(x) &=& - {1 \over 2 \pi i (y-x)}\left(
  {4-q \over 2 (y-x)} \o^{I_1 \cdots I_q}(x) - \partial_x
  \o^{I_1 \cdots I_q}(x) \right) \cr\cr
\o^{I_1 \cdots I_q}(y) \, \psi^C(x) &=& {(-2
i)^{([\frac{q}{2}]-[\frac{q+1}{2}])} \over 2 \pi i (y-x)} \left(-(q-2) 
\partial_x \o^{C I_1 \cdots I_q}(x) + {(3-q) \over (y-x)} \o^{C I_1 
\cdots I_q}(x) \right) \cr\cr
&-& {(-1)^q (i)^{[\frac{q}{2}]-[\frac{q+1}{2}]}\over \pi i(y-x)}
\d^{C I_q}\left( \partial_x \o^{ I_1 \cdots I_{q-1} I}_I(x) + {1 \over
    (y-x)} \o^{ I_1 \cdots I_q I}_I(x) \right) \cr \cr
&-&{i (i)^([\frac{q}{2}] -[\frac{q-1}{2}])\over 2 \pi i(y-x)} {\displaystyle
    \sum_{r=1}^{q-1}} \, \d^{c [I_r }\, \o^{I_1 \cdots {\hat
        {I_r}}\cdots I_q]}(x) \\ \cr
\o^{I_1 \cdots I_q}(y)\,A^{A B}(x) &=&{{1}\over 2\pi i (y-x)} {\displaystyle 
\sum_{r=1}^{q-1}} \{ \o^{ A I_1 \cdots {\hat I_r}\cdots I_q}(x)\,\d^{I_r 
B} -  \o^{ B I_1 \cdots {\hat I_r}\cdots I_q}(x)\,\d^{I_r A}\} \cr\cr
&-& {1 \over 2 \pi i (y-x)} \left(\o^{I_1 \cdots  I_{q-1} A}(x)\,\d^{I_q B} 
+ \o^{I_1 \cdots I_{q-1} B}(x)\,\d^{I_q A}]\right)\\ \cr
\o^{A_1 \cdots A_q}(y)\, \o^{B_1\cdots B_p}(x) &=& { (i)^{[\frac{q}{2}]
  + [\frac{p-2}{2}+2]- [\frac{p+q}2]}\over 2 \pi  (y-x)} {\displaystyle
  \sum_{r=1}^{q}}\, \o^{V_1 \cdots V_{q+r-1} B V_{q+r}\cdots
  v_{q+p-1}}(x)\, \,\times \cr\cr
&\times& \d^{A-1}_{[V_1} \cdots \d^{A_{q-1}}_{V_{q-1}}
\d^{B_1}_{V_q}\cdots \d^{B_r}_{V_{q+r-1}}
\d_B^{A_q}\,\d^{B_{r+1}}_{V_{q+r}}\cdots
\d^{B_{p-1}}_{V_{q+p-2]}}\,\d^{B_p}_{V_{q+p-1}} \cr \cr
&+& {-(-1)^{q (p+q-1)}\over 2 \pi i (y-x)}\, {\displaystyle
  \sum_{r=1}^{q}} \, \o^{V_1 \cdots V_{p+q-3}}(x)\,\,\,\times \cr\cr
&\times& \d^{B_1}_{V_1}\cdots \d^{B_{p-1}}_{V_{p-1}} \d^{{B_p} [
  A_r}\d^{A_1}_{V_p}\cdots
\d^{A_{r-1}}_{V_{p+r-1}}\d^{A_{r+1}}_{V_{p+r}}\cdots\d^{A_q
  ]}_{V_{p+q-3}}\\ \cr
\o^{A_1 \cdots A_q}(y) \, \r^{B_1 \cdots B_p}(x) &=&
-(i)^{[\frac{q}{2}] + [\frac{p}{2}]-[\frac{2q+p-4}{2}]}\left(
{\o^{B_1\cdots B_p A_1\cdots A_q}(x) \over 2\pi i (y-x)^2} + {{\partial_x
  \o^{B_1\cdots B_p A_1\cdots A_q}}(x)\over 2 \pi i (y-x)}\right) \cr
\cr
&+& {\displaystyle \sum_{r=1}^{p+2}}\, {(-1)^{r-1}\over 2 \pi i (y-x)} \,
  \r^{A_1\cdots A_{q-1} B_1\cdots B_r A_q B_{r+1} \cdots B_p}(x)
\end{eqnarray}

\subsection{Short Distance Expansion for $\r^{A_1 \cdots A_q}(y)\,
  O(x)$}
~~~~Our last contribution to the expansions for the coadjoint
representation is given below.  Special cases are explicitly written
and supercede the more general expressions. 
\begin{eqnarray}
R^{\rm J_1 J_2}_{r^{\rm J_1 J_2}} * {\bar U}^{\rm M}_{\o^{\rm M}} &=&
{\bar G}^{\rm J_1}_{(r^{\rm J_1 J_2} \, \o^{\rm M})} \, \d^{\rm M J_2}
  - {\bar G}^{\rm J_2}_{(r^{\rm J_1 J_2}\, \o^{\rm M})} \, \d^{\rm J_1 M}\\ \cr
R^{\rm J_1 J_2}_{(r^{\rm J_1 J_2})} * {\bar U}^{\rm L M N}_{\rm
  \o^{\rm L M N}} &=& \fracm 12 \d^{\rm J_1 L}\, \d^{J_2 M}\, {\bar
  G}^{\rm N}_{(r' \o + r \o')} - 2 \d^{\rm J_2 L}\, \d^{\rm M N}\,
{\bar G}^{\rm J_1}_{(r \o')} + 2 \, \d^{\rm J_1 L} \, \d^{\rm M N} \,
{\bar G}^{\rm J_2}_{(r \, \o')} \\ \cr
{R}^{\rm I_1 \, I_2}_{r^{\rm I_1 I_2}} * {\bar G}^{\rm K}_{\psi^{\rm K}} &=&
\fracm 12 \d^{\rm K [ I_2} \, {\bar U}^{\rm I_1]}_{(r^{\rm I_1 I_2}
  \psi^{\rm K})} - \frac 12 \big( {\bar G}^{\rm J_1}_{(r^{\rm J_1
  J_2}\, \psi^{\rm L})} \, \d^{\rm J_2 \, L} - {\bar G}^{\rm
J_2}_{(r^{\rm J_1 J_2} \psi^{\rm L})} \, \d^{\rm J_1 L}\big) \\ \cr
R^{\rm J_1\cdots J_p}_{r^{\{J_p\}}}* {\bar U}^{\rm {\bar V_1} 
\cdots {\bar V_m}}_{\o^{\{V_m\}}} &=& - \fracm 12 i (i)^{\{[\fracm p2 
] - [\fracm {p+2}2]\}} \, \d_{\rm [{\bar J_1} \cdots {\bar J_p}]}^{\rm 
[ {\bar V_1} \cdots {\bar V_{m-2}} }\,\d^{\rm {\bar V}_{m-1}], {\bar 
V}_{m}}\, \d^{\rm m,p+2}\, {\bar L}_{(r \o)''}\cr \cr
&+&  i^{\{[\fracm p2] -[\fracm{p+1}2]\}} \d^{\rm p+1,m} {\bar G}^{\rm
  \bar V_m}_{(r \o)'}\, \d^{\rm [ {\bar V_1}\cdots {\bar
    V_{m-1}}]}_{\rm [ J_1 \cdots J_p]}  \cr  \cr 
&+& 2 i (-1)^{p+1} \d^{\rm p+3,m} \,(i)^{[\fracm p2] - [\fracm
  {p+2}2]} {\bar G}^{[ {\bar V_1}}_{(r \o)''} \, \d^{\rm {\bar V_2}
  \cdots {\bar V_{m-2}}}_{\rm [ J_1 \cdots J_p]} \,\d^{\rm {\bar
    V_{m-1},\, ]{\bar V_m}} } \cr \cr
&+& i (-1)^p {\bar T}^{\rm {\bar V_1}[ {\bar V_m}}_{(r \o)'} \d^{\rm
  {\bar V_2}\cdots {\bar V_{m-1}}]}_{\rm [ J_1 \cdots J_p ]}
\d^{p+2,m}\cr \cr
&+& (-1)^{\rm pm} (i)^{\{[\fracm m2]+[\fracm p2]-[\fracm{m+p-2}2]
\}} \, {\bar U}^{\rm {\bar V_p+1} \cdots {\bar V_{m-p}}}_{(r \o)'} 
\d^{\rm {\bar V_1}\cdots {\bar V_{p}}}_{\rm [J_1 \cdots J_p]} ~~~,\\
\cr
R^{\rm J_1\cdots J_p}_{r^{\{J_p\}}} * {\bar R}^{\rm {\bar T_1}\cdots
{\bar T_m}}_{\r^{\{T_m\}}} &=&~ \d^{\rm [{\bar T_1} \cdots {\bar T_m}]
}_{\rm [ J_1 \cdots J_p ]}\, \d^{\rm p\,m} \, {\bar L}{}_{(-(\fracm p2 
-2) r' \r - (\fracm p2 - 1)\,r \r')}\cr \cr
&+& (-1)^p\, \{2(i) {\bar G}^{\rm [ {\bar T_1}}_{((2-p) r' \r-(p-1) r  
\r')} \, \d^{\rm {\bar T_2} \cdots {\bar T_m}]}_{\rm [J_1\cdots J_p]} 
\d^m_{p+1}  \cr \cr
&+& (i)(i)^{[\fracm p2]-[\fracm {p-1}2]} \sum_{r=1}^{p} (-1)^r{\bar
  G}^{\rm J_r}_{(r \r)} \d^{\rm m}_{\rm p-1} \d^{[\rm J_1\cdots
  J_{r-1} J_{r+1}\cdots J_p]}_{\rm [ {\bar T_1} \cdots {\bar T_{r-1}}
  {\bar T_{r}}\dots {\bar T_m}]}\} \cr \cr
&+& \sum_{r=1}^{p} (-1)^{r+1} 2 {\bar T}^{\rm [ {\bar T_1}\,|J_r|}_{(r
  \r)} \d^{\rm {\bar T_2} \cdots {\bar T_m} ]}_{\rm [ J_1\cdots
  J_{r-1} \, J_{r+1} \cdots J_p]} \d^p_m \cr \cr
&-& \sum_{r=1}^p (-1)^{r-1} {\bar U}^{\rm [ {\bar T_1} \cdots {\bar
    T_{m-p+1}} \, |J_r|}_{(r \r)} \d^{\rm {\bar T}_{m-p+2}\cdots {\bar
    T}_m ]}_{\rm J_1 \cdots J_{r-1} J_{r+1} \cdots J_{p}} \cr
&+& i^{\{[\fracm p2]+[\fracm{m-p}2]-[\fracm m2]\}} {\bar R}^{\rm [
  {\bar T_{p+1}} \cdots {\bar T_m} }_{(2r' \r + r \r')} \d^{\rm {\bar
    T_1} \cdots {\bar T_p} ]}_{\rm [ J_1 \cdots J_p]}.
\end{eqnarray}

Typical operator product expansions are then ($q > 2$):
\begin{eqnarray}
\r^{A_1\cdots A_q}(y)\, D(x) &=& {(\frac{p}2 -2) \over 2 \pi i
  (y-x)}\left({ \r^{A_1\cdots A_q}(x) \over (y-x)} -(\frac p2 -1)
  \partial_x  \r^{A_1\cdots A_q}(x) \right) \cr\cr
&-&{(i)^{[\frac p2] - [\frac{p+2}2]}\over  2 \pi (y-x)}\left(
  {\o^{A_1\cdots A_q A}_A(x) \over (y-x)^2} + {\partial_x
    \o^{A_1\cdots A_q A}_A(x) \over (y-x)}+{\frac 12 \partial_x^2
    \o^{A_1\cdots A_q A}_A(x)}\right)\\ \cr
\r^{A_1 \cdots A_q }(y) \psi^{B}(x) &=& {i^{[\frac q2] -
    [\frac{q+1}2]}\over 2 \pi i (y-x)}
\left( \partial_x \,\o^{A_1\cdots A_q B}(x) - {1\over (y-x)}
  \o^{A_1\cdots A_q B}(x)\right)\cr \cr
&+&{ 2  (-1)^{q+1} (i)^{[\frac q2 ]- [\frac{q+2}{2} ]}\over  \pi 
  (y-x)}\left( {\o^{B A_1\cdots A_q A}_A(x)\over (y-x)^2} +{\o^{ B
      A_1\cdots A_q A}_A(x)\over (y-x)}
  + \frac 12 \o^{B A_1\cdots A_q A}_A(x) \right)\cr\cr
&+& (-1)^q \left( {(2-p)\r^{B A_1 \cdots A_q}(x) \over \pi (y-x)^2 } -
  {(p-1)\partial_x \r^{B A_1 \cdots A_q}(x) \over \pi (y-x)}\right)
\cr \cr
&+& {(i)^{[\frac q2 ] - [\frac{q-1}2]} \over 2 \pi (y-x)} {\displaystyle
  \sum_{r=1}^q} \d^{A_r B} \, \r^{A_1 \cdots {\hat A_r}\cdots
  A_q}(x)\\\cr
\r^{A_1 \cdots A_q}(y) \, A^{B_1 B_2}(x) &=& (i)^q \left({\o^{[B_1
      |A_1 \cdots A_q | B_2]} \over 2 \pi (y-x)^2} + {\partial_x
    \o^{[B_1 |A_1 \cdots A_q | B_2]} \over 2 \pi (y-x)}\right)\cr \cr
&+ &{\displaystyle \sum_{r=1}^{p}} (-1)^{r+1} {- 1 \over \pi i (y-x)}
  \r^{B_2 A_1 \cdots A_{r-1} B_1 A_{r+1} \cdots A_q} \\\cr
\r^{A_1 \cdots A_q}(y) \, \o^{B_1 \cdots B_n}(x) &=& (-1)^{pn}
(i)^{[\frac{n}2 ] + [ \frac{p}2 ] - [ \frac{n+p-2}2]}\,\left( {\o^{A_1
      \cdots A_q B_1 \cdots B_n}(x) \over 2 \pi i (y-x)^2 } +
  {\partial_x \o^{A_1 \cdots A_q B_1 \cdots B_n}(x) \over 2 \pi i
    (y-x) } \right) \cr\cr
&-& {\displaystyle \sum_{r=1}^p} (-1)^{r-1} {1 \over 2 \pi i (y-x)}
\r^{B_1 \cdots B_{n-r+1} A_r B_{n-r+2} \cdots B_n A_1 \cdots {\hat
    A_r} \cdots A_q}(x) \\ \cr
\r^{A_1 \cdots A_q}(y) \, \r^{B_1 \cdots B_n}(x) &=& i^{[\frac{q}2 ]+
  [\frac{n-2q}{2}] - [ \frac{n-p}2]} \left( {\r^{A_1 \cdots A_q B_1
  \cdots B_n}(x) \over \pi i (y-x)^2} + {\partial_x \r^{A_1 \cdots 
  A_q B_1 \cdots B_n}(x) \over \pi i (y-x)}\right)
\end{eqnarray}

\section{Conclusions}

~~~~This work shows that the dual representation of an algebra provides 
a natural definition of a short distance expansion or operator product 
expansion.  The operators are not built from an enveloping algebra as in 
the Sugawara constructions \cite{Sugawara:1967rw} but instead exists in 
their own right. For example in the section \ref{case1} the operators, 
$D(x)$, satisfy the same operator product expansions as one has in 
Virasoro theories \cite{frenkel} which is built from a lattice.  However, 
notice the $A(x)$ that appears in our discussion of affine Lie algebras 
does not appear in conformal field theory treatments of WZNW
\cite{Wess:yu,Witten:ar,Novikov:ei} where the energy-momentum tensor
is constructed from the current algebra.  Our construction is otherwise 
model independent and can be implemented for  {\em {any}} algebra that 
admits a dual.   Underlying free-field theories are {\em {not}} necessary 
to define these expansions.    Thus, the roles of symmetry groups in 
defining these expansions is moved firmly to the foreground. The origin 
of this model independence is that we exploit the natural bifurcation 
of the initial data where the algebraic elements serve as conjugate 
momenta and the dual of the algebra serves as the conjugate coordinates. 
This feature is shared by the symplectic structures of actions like  
Yang-Mills theory and $N$-extended affirmative actions 
\cite{Gates:2001uu,Boveia:2002gf}.  The coadjoint representation enforces 
a short distance expansion already at the level of the Poisson bracket 
of the elements with no Hamiltonian required.  

Another implication that we wish to note is that since we never relied on 
any specific model in our discussion, this implies that at no point was it 
necessary to invoke Wick rotations to a Euclideanized formulation to justify 
the form of the short-distance expansions.  We note that just as in the case 
of Virasoro theories and some other conformal field theory based operators,
 the central extension can be further restricted by demanding unitarity.  
Choices of the central extension can be determined through the Kac determinant 
\cite{Kac:ge,Friedan:1983xq,Feigin:se}.

This viewpoint  has been applied  to the model-independent $N$-extended 
supersymmetric ${\cal {GR}}$ Virasoro algebra \cite{Gates:1998ss} to obtain 
for the first time its representation in terms of short distance expansions.  
The complete set of such expansions have been presented in the fifth section 
of this work.  This success is expected to open up further avenues of study.  
Since these short distance expansions are now known for arbitrary values of 
$N$, this means that our new results may be used to study the possible existence 
of 1D, $N$ = 16 or $N$ = 32 supersymmetric NSR-type models.  The question of 
whether such an approach can lead to a new manner for probing M-theory 
is now a step closer to being answered.

{\bf {\large {Acknowledgments}}}
~~~~VGJR thanks the Center for String and Particle Theory at the
Physics Department of the University of Maryland for support and 
hospitality.

\end{document}